\documentclass{svjour3}                     

\pdfoutput=1

\smartqed  
\usepackage{graphicx}
\usepackage{rotating}
\usepackage{amssymb}
\usepackage{amsmath}
\usepackage[numbers]{natbib}
\usepackage{bm}
\begin{document}

\title{The dynamics of vortex generation in superfluid $^3$He-B}

\titlerunning{Dynamics of vortex generation}        

\author{R. de Graaf \and R. H\"anninen  \and T.V.~Chagovets \and V.B.~Eltsov
\and M.~Krusius \and R.E. Solntsev}

\authorrunning{R. de Graaf \textit{et al.}} 

\institute{R. de Graaf \and R.E. Solntsev \and V.B. Eltsov \and R.
H\"anninen \and M. Krusius \at Low Temperature Laboratory,
Helsinki University of Technology, FIN-02015 TKK, Finland \\
              Tel.: +358503442189\\
              Fax: +35894512969\\
              \email{rdegraaf@boojum.hut.fi}\\
           \and T.V. Chagovets \at Institute of Physics ACSCR, Na Slovance 2, 182
21 Prague, Czech Republic }

\date{Received: 10 Sep, 2008 / Accepted: 11 Sep, 2008}

\maketitle

\begin{abstract}
A profound change occurs in the stability of quantized vortices in
externally applied flow of superfluid $^3$He-B at temperatures
$\lesssim 0.6\,T_{\rm c}$, owing to the rapidly decreasing damping
in vortex motion with decreasing temperature. At low damping an
evolving vortex may become unstable and generate a new independent
vortex loop. This single-vortex instability is the generic
precursor of turbulence. We investigate the instability with
non-invasive NMR measurements on a rotating cylindrical sample in
the intermediate temperature regime (0.3 -- 0.6)\,$T_{\rm c}$.
From comparisons with numerical calculations we interpret that the
instability occurs at the container wall, when the vortex end
moves along the wall in applied flow.



\end{abstract}

\section{INTRODUCTION}

Turbulence in a rotating fluid is a most frequently encountered
phenomenon, ranging from astrophysical and planetary scales to
engineering problems. Turbulence in a rotating superfluid has not
been extensively investigated, although it is a simpler form of
turbulence from that in viscous fluids and might provide a
shortcut to new understanding. At temperatures below $ 2.5\,$mK,
uniform rotation is one of the few technically feasible means of
generating flow on the global scale. The main reason for studying
superfluid $^3$He-B at these temperatures is to learn about the
influence of a strongly temperature dependent vortex damping
\cite{Turbulence}, or mutual friction dissipation $\alpha (T)$.
With decreasing temperature a sudden onset of turbulence is seen
in applied flow. It is this onset and the  mechanisms behind it
that are described in this report. An instability of a single
vortex evolving in applied flow is at the root of the onset. We
study the instability using a new measuring technique, the
injection of a seed vortex in rotating vortex-free counterflow,
and monitor the subsequent evolution of vorticity with noninvasive
NMR measurement \cite{ROP}.

{\bf Spin up of the superfluid component:}---In uniform rotation
at an angular velocity $\Omega$ the induced applied flow is the
superfluid counterflow (cf) of the normal and superfluid
fractions, the relative velocity $\mathbf{v} = \mathbf{v}_{\rm n}
- \mathbf{v}_{\rm s}$. The viscous normal fraction (with olive
oil-like viscosity) is practically always in solid-body rotation
with the container, $\mathbf{v}_{\rm n} = \mathbf{\Omega} \times
\mathbf{r}$, while the velocity $\mathbf{v}_{\rm s}$ of the
superfluid fraction is produced by the combined flow field from
all quantized vortex lines and from the flow caused by the
boundary conditions on the container walls. Superflow is
characterized by a finite critical velocity $v_{\rm c,exp}$, at
which vortex formation starts. If rotation is started at
temperatures below $T_{\rm c}$, then the vortex-free Landau state
is formed first, where $\mathbf{v}_{\rm s} = 0$ (in the stationary
laboratory frame). When the cf velocity reaches the critical value
$v_{\rm c,exp}$ (somewhere at a rough spot on the container wall),
a vortex is formed, which then evolves from a micron-size loop to
a rectilinear line vortex in the center of the cylinder, aligned
parallel to the rotation axis. Thereby the maximum cf velocity
drops below the critical limit, $v(R) < v_{\rm c,exp}$. If the
external rotation drive $\Omega$ is continuously increased, then
the process repeats and a central cluster of rectilinear line
vortices is formed which is contained and isolated by an annular
layer of vortex-free cf from the cylinder wall. The maximum cf
velocity in the vortex-free flow is at the cylinder wall, with
$v(R) \approx v_{\rm c,exp}$. Within the vortex cluster the global
cf vanishes (on length scales exceeding the inter-vortex distance
$d_{\rm v}$) and $\mathbf{v}_{\rm s}$ is on an average in
co-rotation with $\mathbf{v}_{\rm n}$.

This is the linear well-behaved vortex formation process above
$0.6\,T_{\rm c}$ at high mutual friction damping, where it
provides the ``spin up" of the superfluid component to co-flow
with the normal fraction. In macroscopic flow geometries such
``quasi-intrinsic" vortex formation can be observed in superfluid
$^3$He-B \cite{VorNucleation}, where (depending on surface
roughness of the cylinder wall) the experimental $v_{\rm c,exp}$
is perhaps only an order of magnitude smaller than the theoretical
$v_{\rm c,bulk}$, while in $^4$He-II this has been demonstrated
only with flow through orifices of $\lesssim 1\,\mu$m size.
Typically in a smooth-walled cylinder $v_{\rm c,exp} \sim 1\,$cm/s
at about $0.7\,T_{\rm c}$ from where it decreases approximately as
$v_{\rm c,exp} \propto \sqrt{1-T/T_{\rm c}}$ towards $T_{\rm c}$.
Although this is an order of magnitude smaller than the bulk
liquid critical velocity $v_{\rm c,bulk} \sim 11\,$cm/s (at $P =
29\,$bar liquid pressure), vortex-free cf can thus be maintained
in metastable state up to $v(R) \lesssim v_{\rm c,exp}$, which
substantially influences the NMR absorption response.

\begin{figure}[t]
\begin{center}
\centerline{\includegraphics[width=0.9\linewidth]{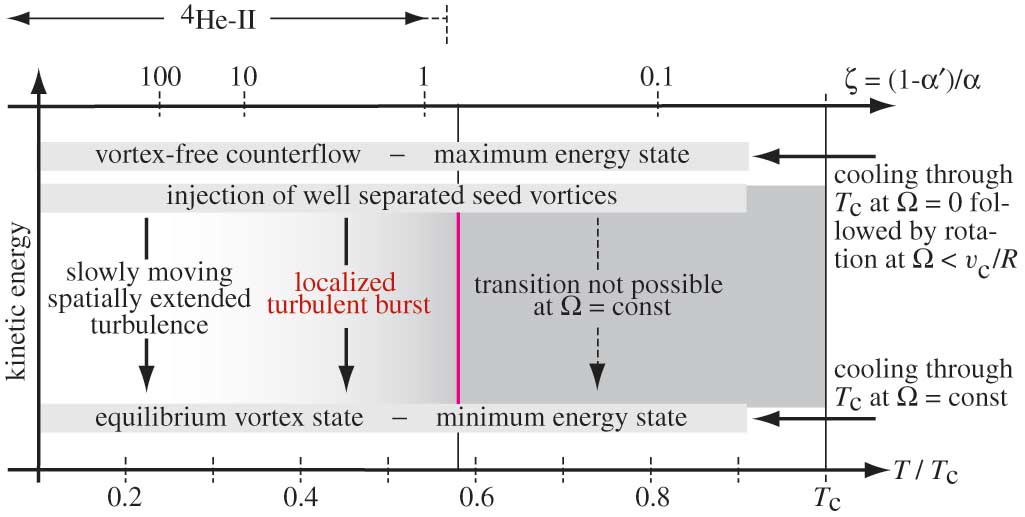}}
\caption{(Color online) Principle of measurements on dynamic
vortex generation. Well separated isolated seed vortices are
introduced in rotating vortex-free cf. The initial high-energy
state of vortex-free flow may then relax to the equilibrium vortex
state via vortex generation processes which become possible at
temperatures below the hydrodynamic transition at $0.59\,T_{\rm
c}$ (at 29\,bar liquid pressure). The instability of a single
evolving seed vortex is the first step. It is followed by a
turbulent burst which is started when the density of evolving
vortices is sufficient. The probability of the combined process
depends on the dynamic mutual friction parameter $\zeta = (1-
\alpha^{\prime})/  \alpha $ which is shown on the top together
with its range of values in superfluid $^4$He.}
\label{RotStates+Injection}
\end{center}
\vspace{-6mm}
\end{figure}

In $^3$He-B mutual friction decreases towards low temperatures
almost exponentially. Thus the transition from the laminar to the
turbulent flow regime at $0.6\,T_{\rm c}$ is abrupt and sudden
\cite{Turbulence}. Below $0.6\,T_{\rm c}$ mutual friction
dissipation becomes sufficiently small so that an evolving vortex
may become unstable and generates a new vortex loop which then
starts expanding independently. This can happen repeatedly at
constant rotation \cite{Precursor}. The experimental signature of
the instability is a sudden burst of turbulence, when the density
of evolving vortices grows sufficiently so that they start
interacting turbulently in the bulk volume. To characterize this
chain of events, we need to know how the first new loop is
generated from a vortex which is evolving in applied flow. This is
investigated here using the scheme outlined in
Fig.~\ref{RotStates+Injection}.

The mechanism by which the superfluid fraction is set into
rotation is often called ``spin up" of the superfluid component
\cite{Andronikashvili}. A sequence of processes is involved in
spin up, one of them being the formation of new vortices. Ideally,
our approach here is to study spin-up at constant rotation
$\Omega$, by injecting a seed vortex in vortex-free flow. On the
macroscopic scale, superfluids often mimic the behavior of viscous
liquids \cite{Vinen}. However, as we shall see below, our
superfluid spin up is rather different from the viscous flow
patterns which evolve in the spin up of classical liquids
\cite{Greenspan}. Owing to its high viscosity, the normal
component is during the spin up in a state of laminar flow. This
is an important simplification over the complicated coupled
turbulent spin up of the normal and superfluid fractions which is
observed in $^4$He-II at higher temperatures \cite{Vinen}.

\begin{figure}[t]
\begin{center}
\centerline{\includegraphics[width=1\linewidth]{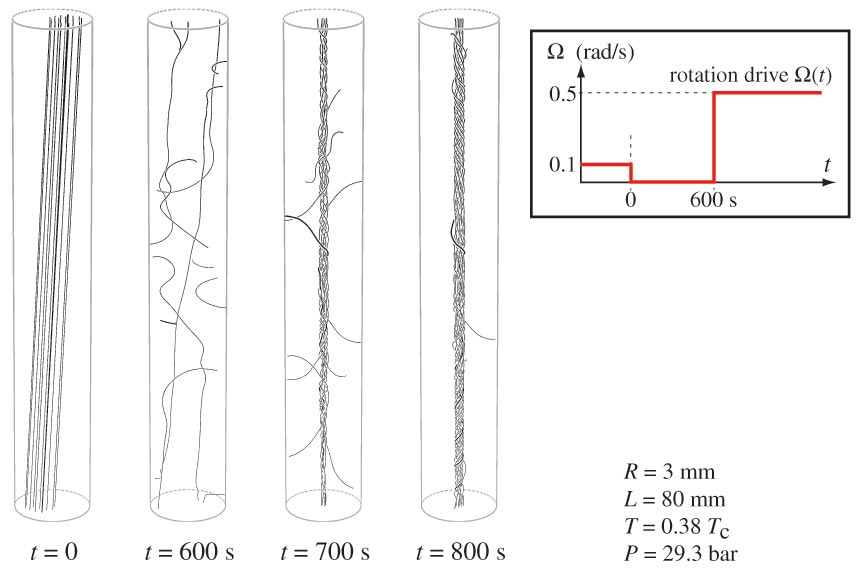}}
\caption{(Color online) Numerical calculation of vortex motions in
a rotating cylinder: $(t \leq 0)$ Initial state with 22 vortices
at 0.1\,rad/s rotation. The vortices have been artificially tilted
by displacing their end points uniformly by 1\,mm at both end
plates of the cylinder, to break cylindrical symmetry. Rotation is
then abruptly reduced to zero, to allow vortices to annihilate.
$(t \leq 600\,$s) After a waiting period $\Delta t = 600\,$s, 12
remanent vortices remain which are here shown at $\Omega =0$.
Rotation is then increased to $\Omega_{\rm f} = 0.5\,$rad/s $(t
\geq 600\,$s) and the 12 remnants start evolving towards
rectilinear lines. This requires that the vortex ends on the
cylindrical wall travel in spiral motion to the respective end
plates. The mutual friction parameters are
$\alpha = 0.18$ and $\alpha^{\prime} = 0.16$ \protect\cite{bevan}.
In the picture the radial lengths have been expanded by two,
compared to axial distances.} \label{RemEvolution}
\end{center}
\vspace{-6mm}
\end{figure}

\begin{figure}[t]
\begin{center}
\centerline{\includegraphics[width=0.9\linewidth]{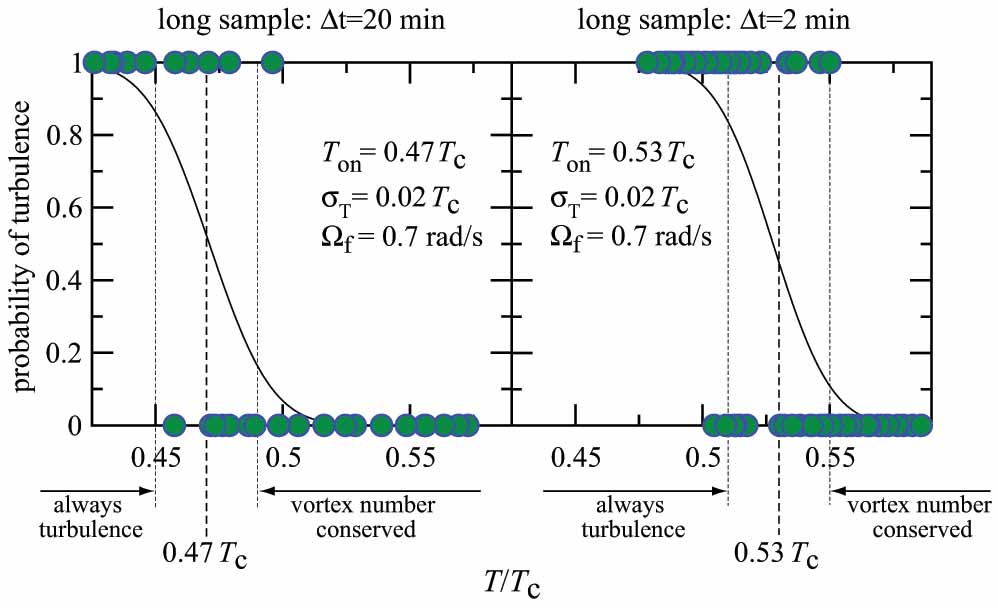}}
\caption{(Color online) Measurements on the onset temperature
$T_{\rm on}$ of turbulence. The measurement starts from an initial
state which is obtained by decelerating an equilibrium vortex
state at 1.7\,rad/s to $\Omega = 0$ at a rate 0.01\,rad/s$^2$. The
remaining vortices are left to annihilate \cite{DynamicRemnants}
for a period $\Delta t$ at $\Omega = 0$. Rotation is then
increased to $\Omega_{\rm f}$ at a rate 0.02\,rad/s$^2$. When all
transients have decayed the number of vortices is measured in the
final steady state at $\Omega_{\rm f}$. The result is plotted as a
function of temperature with 30 -- 40 data points per panel.  The
solid curve is a gaussian fit which represents the probability for
turbulence with a half width $\sigma_{\rm T} = 0.02\,T_{\rm c}$
and centered around $T_{\rm on}$. Comparing results in the two
panels for $\Delta t = 20\,$min and 2\,min, we see that $T_{\rm
on}$ decreases with increasing $\Delta t$, since the number,
average size, and density of remnants is reduced as $\Delta t$
increases.  Parameters: $R = 3\,$mm, $L = 110\,$mm, and $P =
29.0\,$bar. } \label{RemOnsetTempLongCyl}
\end{center}
\vspace{-6mm}
\end{figure}

{\bf Generic properties of single-vortex instability:}---The
single-vortex instability below $0.6\,T_{\rm c}$ and the phenomena
which it starts in a rotating column with circular cross section
have been described in Ref.~\cite{Eltsov:2008}. Here we repeat
some central features in the light of Figs.~\ref{RemEvolution} and
\ref{RemOnsetTempLongCyl}, before we turn to a closer
characterization of the instability itself. One type of
measurement on evolving vortices in a rotating cylinder is
illustrated in Fig.~\ref{RemEvolution}. This numerical calculation
with the vortex filament method \cite{Simulation} (see
Sec.~\ref{VortexFilamentCalculation}) monitors remanent vortices
\cite{DynamicRemnants} in a rotating cylinder (with radius $R$ and
length $L$ at constant $\Omega$), in a situation where the vortex
instability does not occur. The purpose is to illustrate the
motion of remanent vortices while they evolve from short curved
vortices to rectilinear lines. The practical outcome from such a
measurement is analyzed in Fig.~\ref{RemOnsetTempLongCyl} as a
function of temperature.

A measurement of the kind depicted in Fig.~\ref{RemEvolution}
proceeds as follows: As shown in the inset, an equilibrium vortex
state is first decelerated to zero rotation. After a waiting time
$\Delta t$ at zero rotation, some vortices have not yet managed to
annihilate. Their annihilation time is governed by mutual friction
damping $\alpha (T)$ and increases rapidly with decreasing
temperature. When rotation is next suddenly increased from zero to
a steady value $\Omega_{\rm f}$, the remaining vortices start
expanding towards their stable state as rectilinear lines. This
motion proceeds such that the vortex ends on the cylindrical wall
travel along a spiral trajectory. During such helical motion on
outer evolving vortex is wound around the straighter vortices in
the center \cite{VorInstability}. The spiral motion is evident on
the far right where we see a cluster of helically twisted vortices
in the center of the cylinder. This state is still evolving, since
ultimately also the helical twist relaxes to rectilinear lines,
when the vortex ends slowly slide along the end plates of the
cylindrical container \cite{TwistedVorState}.

Using the measuring routine in Fig.~\ref{RemEvolution}, we can
find the onset temperature of the single vortex instability. This
is done by measuring the number of rectilinear vortex lines in the
final state at different temperatures. A rough value for the onset
temperature can be located easily: (1) Well above $T_{\rm on}$ the
number of vortices does not change during the evolution of the
remnants and thus in the final state there are only few vortices.
(2) Well below $T_{\rm on}$, in contrast, the instability occurs
always and is followed by a turbulent burst which starts the
evolution towards the equilibrium vortex state \cite{Eltsov:2008}.
These two types of final states correspond to very different NMR
absorption spectra. In Fig.~\ref{RemOnsetTempLongCyl} the
distribution of these two final states is plotted in the immediate
vicinity of $T_{\rm on}$. Each data point corresponds to an
independent measurement of the final state, after the remanent
vortices have expanded in the applied cf at constant external
conditions ($\Omega_{\rm f}$, $T$, and $P$) and all transients
have relaxed. Only in the temperature regime $T \sim T_{\rm on}$
the final state is unpredictable. By fitting the measurements to
the normal distribution we get the probability of starting the
turbulent burst, the quantity plotted on the vertical scale in
Fig.~\ref{RemOnsetTempLongCyl}.

Two cases are compared in Fig.~\ref{RemOnsetTempLongCyl}: on the
right the waiting time at zero rotation is $\Delta t = 2\,$min,
while on the left it is $\Delta t = 20\,$min. This turns out to
yield different values for $T_{\rm on}$. With $\Delta t = 2\,$min,
the number of remanent vortices $\cal{N}_{\rm i}$ is about ten
times larger (approximately 60 at about $0.53\,T_{\rm c}$) than
with $\Delta t = 20\,$min (approximately 10 at about $0.47\,T_{\rm
c}$) \cite{DynamicRemnants}. Because of the lower seed vortex
density in the latter case, the onset moves from $0.53\,T_{\rm c}$
to $0.47\,T_{\rm c}$. Similarly, if $\Omega_{\rm f}$ is reduced
from its value of 0.7\,rad/s in Fig.~\ref{RemOnsetTempLongCyl},
the onset moves to a lower temperature. This means that, to start
turbulence, a sufficiently low mutual friction dissipation $\alpha
(T)$ is the most important condition, but what also matters is the
velocity of the applied cf and the initial number, configuration,
and density of seed vortices.

The measurements in Fig.~\ref{RemOnsetTempLongCyl} are not
conducted with single-vortex resolution: The number of rectilinear
vortices $N$ in the final state is determined within $\pm 10\,$
lines. While the number of seed vortices is small, in the
equilibrium vortex state it is $N_{\rm eq} \approx 640$, and thus
these two situations are easily distinguished. In the onset regime
$T \sim T_{\rm on}$, both failed and successful attempts for a
transition to turbulence occur. In the failed attempts no increase
in the number of vortices is measured, which means that only few
new vortices are needed before turbulence manages to switch on.
Secondly, all final states in Fig.~\ref{RemOnsetTempLongCyl} are
either equilibrium vortex states or states with essentially no new
vortices. This suggests that when turbulence is switched on, a
surplus of vortices is created in the turbulent burst.
Simultaneously the polarization of the vortices along the rotation
axis grows to $\gtrsim 90\,$\% and the number of vortices adjusts
itself approximately to that of the equilibrium vortex state.

It needs to be pointed out that the generation of new vortices in
Fig.~\ref{RemOnsetTempLongCyl} is very different from regular
vortex formation at temperatures above $0.6\,T_{\rm c}$. The
vortices in Fig.~\ref{RemOnsetTempLongCyl} are created at constant
rotation $\Omega_{\rm f}$ at a low cf flow velocity $\lesssim
2\,$mm/s relatively evenly along the entire column
\cite{Precursor}. If turbulence is started, then the macroscopic
cf velocity ultimately drops close to zero. This means that vortex
generation proceeds until completion (with the exception of a few
rare examples which have been observed only in the onset regime,
$T \approx T_{\rm on}$). In contrast, in regular vortex formation
above $0.6\,T_{\rm c}$ at a rough spot on the cylinder wall (or at
a wall defect called a ``vortex mill" \cite{VortexMill}), the flow
becomes sub-critical after the first new vortex. A second vortex
is not created (in the ideal situation \cite{VorNucleation}),
unless $\Omega$ is again increased.

Thus to summarize, in contrast to the quasi-intrinsic spin-up
process above $0.6\,T_{\rm c}$, the single-vortex instability as
the precursor to turbulence has unmistakable features, which
become evident when one cools down into the intermediate
temperature regime, from 0.6 to 0.3\,$T_{\rm c}$. Here the
quasiparticle mean free path $\ell \lesssim 50\mu{\rm m}$ is still
smaller than the typical inter-vortex distance $d_{\rm v} \sim
0.2\,$mm and much less than the sample size $R = 3\,$mm. In the
following we focus on the single-vortex instability and the
immediately following turbulent burst.

\section{MEASUREMENTS ON SINGLE-VORTEX INSTABILITY}
\label{Measurements}

{\bf Rotating flow states:}---Controlled seed vortex injection and
the calibration of measured NMR signals requires stability and
reproducibility of different rotating flow states. Above
$0.6\,T_{\rm c}$ one can experimentally prepare a state with any
number of rectilinear vortex lines in the central cluster up to
the equilibrium number: $N \leq N_{\rm eq}$. These are called (i)
the vortex-free state $(N=0)$, (ii) a metastable vortex cluster
$(N < N_{\rm eq})$, and (iii) the equilibrium vortex state $N
\approx N_{\rm eq}$.  When a vortex is formed as a small loop, it
expands in spiral motion to a rectilinear line and becomes part of
the central cluster, as seen in Fig.~\ref{RemEvolution}. Thereby
the radius $R_{\rm o}$ of the cluster increases and the cf
velocity is reduced: $v = \Omega r - \kappa N/(2\pi r)$. Here
$\kappa = h/(2 m_3)$ is the superfluid circulation quantum, $N$ is
the number of vortices in the cluster, and $R_{\rm o} < r < R$.
The cluster reaches its maximum radius in the equilibrium vortex
state, where $R_{\rm o} = R - d_{\rm eq}$ and $d_{\rm eq} \gtrsim
d_{\rm v} \approx \sqrt{\kappa/(2\Omega)}$ is the equilibrium
width of the vortex-free annulus between the cluster and the
cylinder wall. Independently of the number of vortices $N$, the
maximum applied flow is at the cylindrical boundary. At higher
$\Omega$ (when there is no annihilation barrier
\cite{TiltedContainer}) the equilibrium vortex state is also the
state with the maximum number of vortices (at a given value of
$\Omega$ in stable conditions).

The precondition for generating these rotating states is a
sufficiently high and stable critical velocity of vortex formation
$v_{\rm c,exp}$ \cite{VorNucleation}. Since surface roughness
reduces $v_{\rm c,exp}$, smooth and clean container walls are
important. Thus $v_{\rm c,exp}$ is container dependent and varies
even from one cool down to the next, presumably owing to frozen
residual gas crystallites on the walls. Since the vortex number is
reliably conserved during vortex formation processes only above
$0.6\,T_{\rm c}$, a particular rotating state generally has to be
formed at high temperatures, but can then be cooled to low
temperatures at constant rotation. In particular, to have a
rotating vortex-free sample at low temperatures, the procedure is
to warm up first above $0.6\,T_{\rm c}$, where the annihilation of
remnants is rapid \cite{DynamicRemnants}. The sample is kept there
at zero rotation for 10 to 20\,min and is then cooled down in
rotation (with $\Omega < v_{\rm c,exp} / R$).

\begin{figure}[t]
\begin{center}
\centerline{\includegraphics[width=0.3\linewidth]{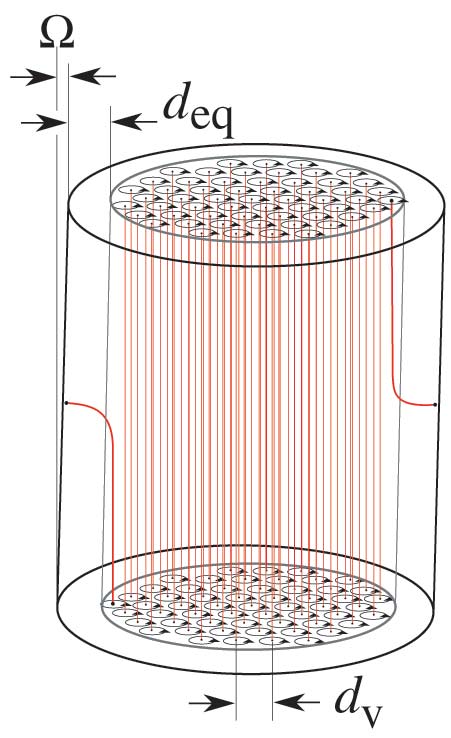}}
\caption{(Color online) Sketch of equilibrium vortex state in a
slightly tilted cylinder with two vortices connecting to the
cylindrical side wall. }\label{TiltedEquilVorState}
\end{center}
\vspace{-8mm}
\end{figure}

The equilibrium vortex state is a particular case. Ideally, when
the sample and rotation axes are perfectly aligned, all vortices
in the equilibrium cluster are rectilinear. In practice, some
misalignment is unavoidable, typically of order $\sim 1^{\circ}$.
The result of this is that some of the outermost vortices are then
curved and attached at one end to the cylindrical wall
(Fig.~\ref{TiltedEquilVorState}). When $\Omega$ is increased,
these vortices start to spiral towards rectilinear lines. Below
$0.6\,T_{\rm c}$ this evolution may become unstable and leads to a
turbulent burst, as discussed in the context of
Fig.~\ref{RemOnsetTempLongCyl}. Thus the equilibrium vortex state
can be used for measurements on the single-vortex instability in a
similar manner as remanent vortices.

{\bf Seed vortex injection:}---By introducing seed vortices in
applied flow by externally controlled means, we can monitor vortex
evolution as a function of time at constant rotation. In rotating
flow any vortex which is not rectilinear can be used as seed
vortex, {\it i.e.} a small vortex loop or any section of a longer
vortex whose configuration changes appreciably as a function of
time. The active section of the seed vortex is outside the vortex
cluster in the counterflow region, where the macroscopic cf
velocity $v(r) \neq 0$, and is curved, with one or both ends
connected to the cylindrical side wall. A number of different
methods exist to start seed vortex evolution \cite{Injection,ROP}.
One example is the use of remanent vortices
(Figs.~\ref{RemEvolution} and \ref{RemOnsetTempLongCyl}) and a
second the curved peripheral vortices of the equilibrium vortex
state (Fig.~\ref{TiltedEquilVorState}).

An efficient injection method is based on the superfluid
Kelvin-Helmholtz instability \cite{KH-Instability} of the phase
boundary between the A and B phases of superfluid $^3$He. In this
instability a tightly packed bundle of several approximately
parallel vortex loops escapes across the AB interface into the
vortex-free B-phase flow \cite{KH-Injection}. This allows
immediate inter-vortex interactions and starts the turbulent
burst. There are several further important points to note. First,
the burst follows instantaneously the KH instability and no
precursory vortex generation via the single vortex instability is
observed before the burst. Secondly, the burst can be localized to
the immediate vicinity of the AB phase boundary \cite{ROP}.
Thirdly, the KH transition to turbulence is independent of the
applied flow velocity $v(\Omega,N,R)$ \cite{Turbulence}, unlike
other turbulent bursts which require the single-vortex instability
as precursor \cite{TurbTrans}. Most importantly however, compared
to other injection methods,  KH injection gives the highest onset
temperature for turbulence. This means that the turbulent burst
can happen at a higher value of vortex damping $\alpha (T)$ than
where the single-vortex instability becomes possible. In KH
injection the transition temperature to turbulence displays a
typical narrow normal distribution \cite{TurbTrans} which is
similar to that in Fig.~\ref{RemOnsetTempLongCyl} (except for the
value of $T_{\rm on}$ which is higher). This similarity suggests
that plots of the transition to turbulence, like that in
Fig.~\ref{RemOnsetTempLongCyl}, describe the transition
probability in a situation when enough vortices have already been
created by the precursor mechanism so that turbulence can switch
on. For these reasons KH injection is at present the process which
is believed to identify most clearly the hydrodynamic phase
transition between regular and turbulent vortex dynamics.

Finally, we note that the capture reaction of a thermal neutron in
$^3$He-B releases vortex rings in a predictable manner into
vortex-free cf \cite{NeutronReview}. Vortex formation is here
explained to happen via the Kibble-Zurek mechanism \cite{Zurek},
during a sudden temperature quench from the normal phase into the
superfluid. This process can be reliably adjusted to emit only one
single vortex ring into vortex-free B-phase flow. The telltale
observation is that even this single seed vortex ring manages to
start the turbulent burst at sufficiently low mutual friction
dissipation \cite{NeutronInjection}.

\begin{figure}[t]
\begin{center}
\centerline{\includegraphics[width=0.45\linewidth]{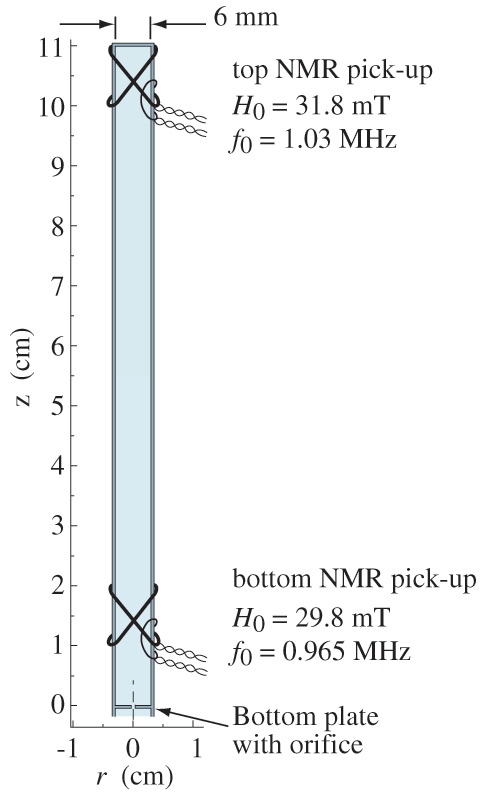}}
\caption{(Color online) $^3$He sample setup. The oval $2 \times
15$-turn superconducting  NMR tank circuit coils are connected
inductively via a two-turn pick-up loop to a room temperature
preamplifier. This provides weak coupling to the pick-up coil, so
that the $LC$ resonator is not excessively loaded and a Q-value of
$\sim 6000$ is achieved. Two solenoidal superconducting magnets
\cite{NeutronInjection} (which are not shown) provide the
homogeneous axially oriented polarizing fields for NMR.
}\label{setup}
\end{center}
\vspace{-8mm}
\end{figure}

For the present purpose, to investigate the single-vortex
instability, we need an injection technique where the seed
vortices are initially far apart at low applied cf velocity. We
can use either remanent vortices \cite{DynamicRemnants} or the
curved peripheral vortices of the equilibrium vortex state, when
the sample and rotation axes are not perfectly aligned
\cite{TiltedContainer}. In both cases rotation is increased
rapidly from the initial state at $\Omega_{\rm i}$ to a final
constant value $\Omega_{\rm f}$ where the evolution is recorded at
constant external conditions. For remanent vortices the initial
state is at zero rotation, $\Omega_{\rm i} =0$, while in the case
of the equilibrium vortex state $\Omega_{\rm i}$ has some constant
low value. In a strict sense these two methods do not represent
injection into flow at constant cf velocity, like the
Kelvin-Helmholtz instability or the Kibble-Zurek mechanism in
neutron irradiation. However, in practice they achieve the same
result, namely placing evolving vortices in rotating cf. In the
onset regime, $0.3\,T_{\rm c} < T \leq 0.6\,T_{\rm c}$, the
probability of turbulence depends primarily on the final rotation
velocity $\Omega_{\rm f}$ and only weakly on the acceleration
$\dot{\Omega}$ used to reach $\Omega_{\rm f}$. We use
$\dot{\Omega} \sim 0.02\,$rad/s$^2$, which in practice mimics a
step increase to $\Omega_{\rm f}$.

{\bf Experimental setup:}---The measurements are performed in a
rotating nuclear demagnetization cryostat in which the liquid
$^3$He sample can be cooled below $0.2\,T_{\rm c}$ in rotation up
to 3\,rad/s. The temperature is determined from the frequency
shifts in the NMR spectra \cite{Hakonen,Ahonen} and below
$0.3\,T_{\rm c}$ from the damping of a quartz tuning fork
oscillator \cite{Fork}. The sample container (Fig.~\ref{setup}) is
a fuzed quartz tube of radius $R=3$\,mm and length $L=110\,$mm,
filled with liquid $^3$He at a pressure of $P = 29\,$bar. An
aperture of 0.75\,mm diameter in the center of the bottom end
plate restricts the flow of vortices into the sample from the heat
exchanger volume below.

\begin{figure}[t]
\begin{center}
\centerline{\includegraphics[width=0.85\linewidth]{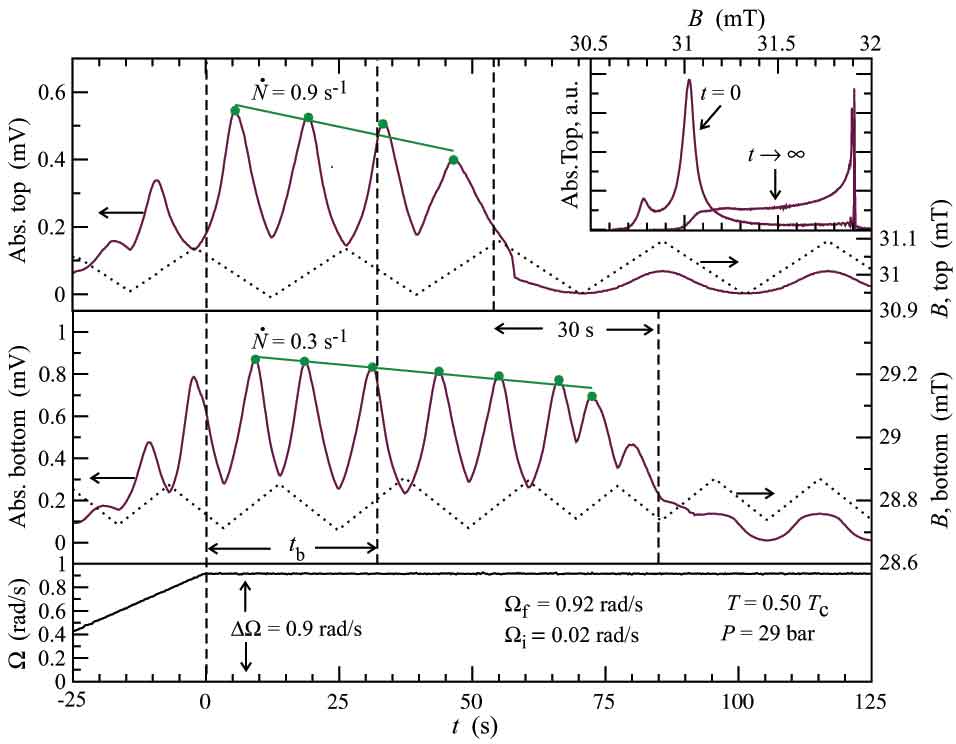}}
\caption{(Color online) NMR record of vortex generation in applied
cf. {\it (Insert)} NMR absorption line shapes at $0.51 \, T_{\rm
c}$ and $\Omega_{\rm f} = 0.92\,$rad/s. The two line shapes
represent: (i) The state after increasing rotation from the
initial equilibrium vortex state at $\Omega_{\rm i}= 0.02\,$rad/s
to $\Omega_{\rm f}$. Here the total number of vortices is still
close to the initial value $N_{\rm i} \approx 5$. This state is
distinguished by the large shifted \textit{cf peak} on the left.
(ii) Final equilibrium vortex state with $N_{\rm eq} \approx 840$
(at $t > 500\,$s), which is marked by increased absorption
bordering to the Larmor edge on the right. Both line shapes have
been measured with the top detector. {\it (Main Panel)} The
\textit{cf peak height} as a function of time after increasing
rotation to $\Omega_{\rm f}$ at $0.50 \, T_{\rm c}$ (which is
below $T_{\rm on} = (0.54 \pm 0.02)\, T_{\rm c}$). The reduction
in peak height represents the gradual increase in the number of
vortex lines in the central cluster. The outputs from the top and
bottom detectors are shown. These are not identical since the
vortex formation rates $\dot{N}$ and the moments in time when the
peak heights collapse may differ along the sample. The sawtooth
waves represent the NMR field sweeps around the location of the cf
peak which is shifted far from the Larmor value. {\it (Bottom)}
Rotation drive $\Omega (t)$ as a function of time. Time $t=0$ is
placed at the point where the final rotation velocity $\Omega_{\rm
f}$ is reached.} \label{WallMediatedInstability}
\end{center}
\vspace{-6mm}
\end{figure}

The continuous-wave NMR absorption line shape (see insert in
Fig.~\ref{WallMediatedInstability}) is measured non-invasively
with two detector coils at both ends of the sample cylinder. The
signal is recorded with constant frequency excitation by sweeping
the polarizing magnetic field. The number of vortex lines $N$ in
the central vortex cluster is obtained from the NMR line shape
either experimentally, by comparing to a reference spectrum which
has been formed with a known number of vortices
\cite{NeutronInjection}, or theoretically, by comparing to a
calculated reference, obtained from calculations of the order
parameter texture \cite{jKopu}. The latter method is applied in
this work with an overall accuracy which we estimate to be within
$\pm 30\,$\%. The calibrations are affected by the misalignment
between the rotation and sample axes. It was measured to be
$0.64^{\circ}$. In the following we make use of this experimental
artifact which breaks cylindrical symmetry and makes the
equilibrium vortex state at some low initial rotation velocity
$\Omega_{\rm i} \neq 0$ useful as a reproducible source of seed
vortices.

Such measurements are possible only if the critical velocity
$v_{\rm c,exp}(T,P)$ is stable and well-behaved. Since surface
defects and dirt on the cylinder wall act as sites for nucleation,
pinning, and even trapping of vortices, the quartz walls are
carefully etched and cleaned. In spite of this some variation in
critical velocity is observed from one cool down to the next,
indicating that perhaps frozen gas particles are involved. The
sample container in Fig.~\ref{setup} has been in continuous use
since a few years, with occasional warm ups to liquid nitrogen
temperatures to clean the dilution refrigerator circulation from
air plugs or to room temperature to modify the experiment. During
the last 12 months $\sim 90\,$\% of cool downs at 0.8\,rad/s to
below $0.20\,T_{\rm c}$ remain vortex-free, while at 0.9\,rad/s
only $\sim 20\,$\% of such attempts are successful. Before that,
the same container could be regularly cooled down in the
vortex-free state at 1.2\,rad/s. The exclusion of isolated surface
defects which trap vortices is no simple task for a large cylinder
like that in Fig.~\ref{setup}. Note that below $0.3\,T_{\rm c}$
even a single expanding remnant may start a turbulent burst and
will then transfer the sample to the equilibrium vortex state.
Probably the maximum vortex-free flow is here limited by isolated
bad spots of dirt or defects on the cylindrical wall where
vortices can be trapped as small loops, while they try to
annihilate in zero flow. A trapped loop has a certain critical cf
velocity at which it can start expanding. This velocity depends on
the radius of the loop and its orientation. Thus the trapping site
controls the flow velocity where the first trapped remnant starts
evolving. During annihilation the isolated traps are randomly
loaded with a remanent loop and thus the critical velocity varies
from one measurement to the next.

\begin{figure}[t]
\begin{center}
\centerline{\includegraphics[width=0.7\linewidth]{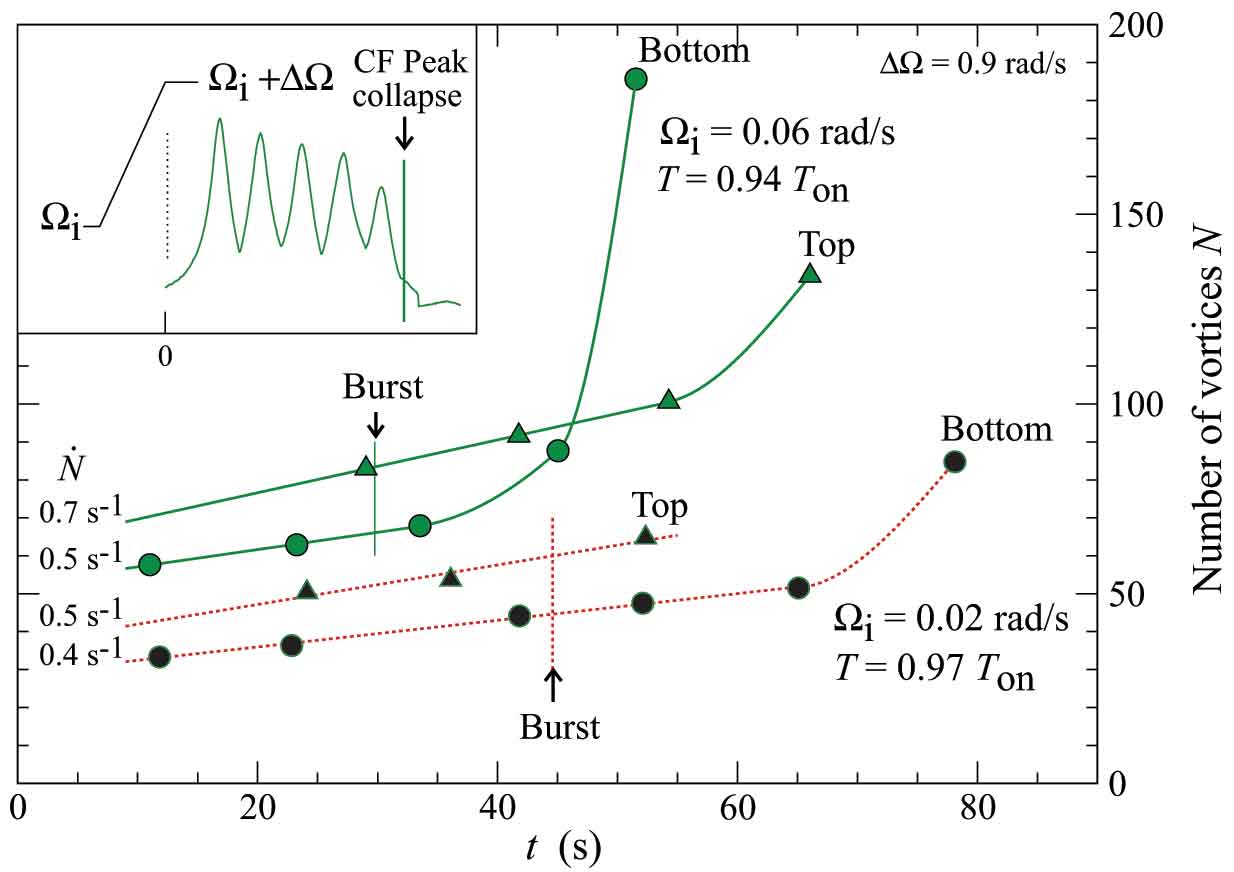}}
\caption{(Color online) Number of vortex lines $N(t)$ in central
vortex cluster as a function of time during continuous vortex
formation owing to the single-vortex instability. Two measurements
are shown which start from equilibrium vortex states at
$\Omega_{\rm i} = 0.02\,$ and 0.06\,rad/s. The calculated total
number of vortices $ N_{\rm i}$ is initially approximately 5 and
33, respectively, while the number of seed vortices $\cal{N}_{\rm
i}$, which connect to the cylindrical wall, is 3 -- 4 in the
former and about 20 in the latter case. Owing to difficulties in
initial equilibration and the presence of a finite annihilation
barrier \cite{TiltedContainer} at $\Omega_{\rm i} = 0.02\,$rad/s,
the extrapolated number of initial vortices appears to be 3 -- 4
times larger than the calculated estimate at very low velocities
like 0.02\,rad/s. The vertical arrows indicate when the turbulent
burst occurs in these two examples. As in
Fig.~\ref{WallMediatedInstability}, the burst does not take place
inside either of the detector coils, but somewhere else in the
long column. Since $T_{\rm on}$ increases with $\cal{N}_{\rm i}$,
$T_{\rm on}$ has a different value for the two cases in this
figure: it is $0.538\,T_{\rm c}$ for $\Omega_{\rm i} =
0.02\,$rad/s, while it is $0.551\,T_{\rm c}$ for $\Omega_{\rm i} =
0.06\,$rad/s. {\it (Insert)} Rotation drive $\Omega (t)$ and cf
peak height measured with the bottom coil at $0.94 \, T_{\rm on}$.
Note that Figs.~\ref{RemOnsetTempLongCyl},
\ref{WallMediatedInstability}, \ref{VorInstability}, and
\ref{BurstTime} -- \ref{BurstLocation} illustrate the
single-vortex instability in a long column (with length/diameter
$\sim 20$), which is sampled with two detectors, to record the
time evolution simultaneously in two places.
Figs.~\ref{PrecursorSlope}, \ref{NdotHistogram}, and
\protect\ref{TopBottomComparisonTon} in turn, describe the
situation when the column is divided with an A-phase separation
layer \cite{KH-Instability} in two independent samples with
length/diameter $\sim 10$, so that the initial equilibration time
to the equilibrium vortex state at $\Omega_{\rm i}$ becomes
shorter. } \label{VorInstability}
\end{center}
\vspace{-8mm}
\end{figure}

{\bf Measuring procedure:}---In the present work the evolving seed
vortices are either remanent vortices (see
Fig.~\ref{RemEvolution}) or vortices curving to the cylindrical
side wall in the equilibrium vortex state (see
Fig.~\ref{TiltedEquilVorState}). Accordingly, no long temperature
sweeps are needed here. The rotation drive $\Omega$ is simply
changed at constant temperature according to a protocol which for
remanent vortices is shown in the inset of
Fig.~\ref{RemEvolution}. The initial state is first formed by
decelerating from high rotation with a large number of vortices to
some low rotation velocity $\Omega_{\rm i}$, where rotation is
maintained constant for a period $\Delta t$. If $\Omega_{\rm i} =
0$, then the waiting period $\Delta t$ at zero applied flow
controls the number of remanent vortices \cite{DynamicRemnants},
as seen in Fig.~\ref{RemOnsetTempLongCyl}. If $\Omega_{\rm i} \neq
0$, then we generally choose $\Delta t = 300\,$s, which allows the
vortex array to approach closer to its equilibrium state. To start
the single-vortex instability, rotation is next increased by a
fixed increment $\Delta \Omega$ at $\dot{\Omega} \sim
0.02\,$rad/s$^2$ to $\Omega_{\rm f}$, where it is kept constant
and the evolution is recorded. In
Fig.~\ref{WallMediatedInstability} the NMR response is shown for
an example case in the onset regime $ T \approx T_{\rm on}$, where
vortex generation starts spontaneously and is finally terminated
in a turbulent burst.

The global superfluid cf from the rotation increase to
$\Omega_{\rm f}$ produces a large absorption peak in the NMR
spectrum which is shifted far from the Larmor resonance. In the
main panel of Fig.~\ref{WallMediatedInstability} the height of
this cf peak is monitored. The reduction in peak height  as a
function of time measures the increase in the number of vortices
$N$ in the central cluster. Well above $T_{\rm on}$ the cf peak
height remains constant, as no new vortices are generated, but in
the onset regime $T \sim T_{\rm on}$ the height may decrease
continuously, as seen in this example in
Fig.~\ref{WallMediatedInstability}. The initial slow rate of
height reduction we attribute to vortex generation by the
single-vortex instability. According to the calculated
calibrations of the cf peak heights, the measured $\dot{N}$
corresponds to adding a rectilinear vortex line every few seconds
to the central cluster. The final sudden collapse in height (after
about 50\,s in the top and 85\,s in the bottom detector) marks the
arrival of the equilibrium number of vortices to the respective
detector coil.

\begin{figure}[t]
\begin{center}
\centerline{\includegraphics[angle=0,width=0.7\linewidth]{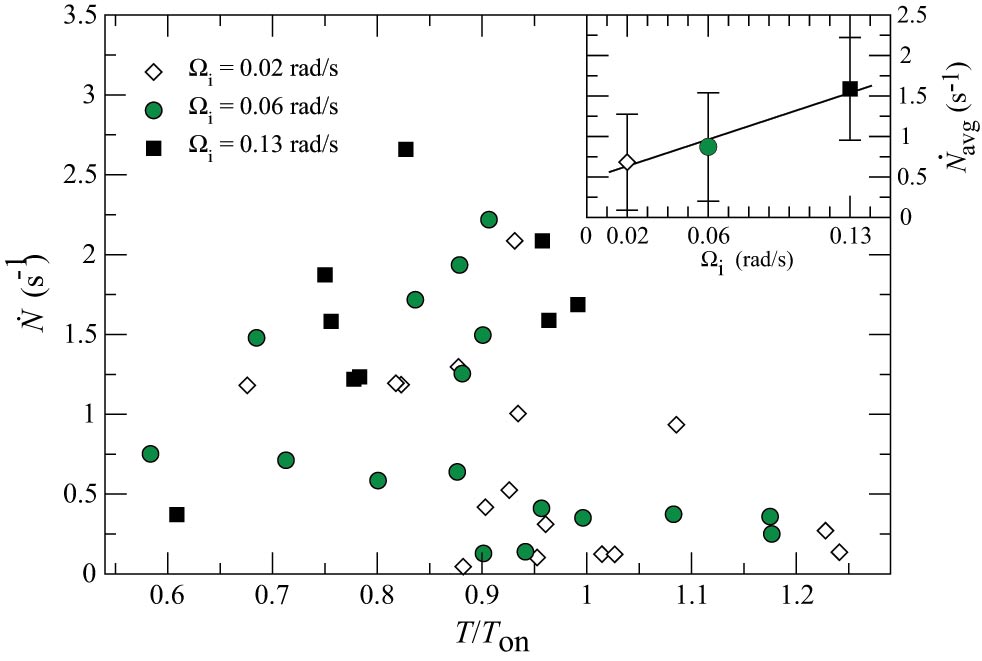}}
\caption{(Color online) Initial rate of vortex generation,
$\dot{N}$ (at $t=0$), as a function of normalized temperature
$T/T_{\rm on}$ for events with long precursory vortex generation
($t_{\rm b} \gtrsim 20\,$s). These measurements are performed with
the sample divided in independent top and bottom sections,
separated by a narrow layer of $^3$He-A as barrier \cite{ROP}.
Initially both sections are in the equilibrium vortex state at
$\Omega_{\rm i}$. To start the generation of new vortices, a rapid
increase in rotation by $\Delta \Omega = 0.7\,$rad/s is applied.
The plot shows that at the final velocity $\Omega_{\rm f} =
\Omega_{\rm i} + \Delta \Omega$ events with a measurable slow
$\dot{N}$ are all in the onset regime $T \approx T_{\rm on}$: at
high temperatures $T / T_{\rm on} > 1.2$ no cases of dynamic
vortex generation were observed, while at low temperatures $T /
T_{\rm on} < 0.6$ all measurements ended in a turbulent burst
which developed too fast to provide a measurement of $\dot{N}$.
{\it (Insert)} The average $\dot{N}$ of the data in the main panel
plotted as a function of $\Omega_{\rm i}$. Thus $\dot{N}$
increases approximately linearly with $\Omega_{\rm i}$, in other
words with the initial number of seed vortices $\cal{N}_{\rm i}$.
The seed vortices are here the curved peripheral vortices of the
equilibrium vortex state which at one end connect to the
cylindrical side wall. As before, the measured value of $T_{\rm
on}$ is different for each of the three values of $\Omega_{\rm i}$
in this figure. Also, vortex generation by the single-vortex
instability is a stochastic event; this is the origin for the
scatter. } \label{PrecursorSlope}
\end{center}
\vspace{-8mm}
\end{figure}

The collapse of the cf peak is the signal that the turbulent burst
has occurred. From the site of the burst a vortex front propagates
both up and down along the rotating column \cite{Eltsov:2008}.
When the front passes through a detector coil, the cf peak height
drops to zero. Above $0.4\,T_{\rm c}$ the longitudinal propagation
velocity $V_{{\rm F}}$ of the front is approximately the same as
that of the end point of a single vortex while it spirals along
the cylindrical wall \cite{PropagationVelocity},  $v_{{\rm L}z}
\approx \alpha \Omega R$  (in an originally vortex-free rotating
column).  Recently measurements \cite{Front} on the front velocity
were extended to temperatures below $0.2\,T_{\rm c}$. Using these
later values of $V_{{\rm F}} = \alpha_{\rm eff} \Omega R$ and
correcting them for the momentary number of vortex lines $N$ in
the central cluster around which the front spirals, $V_{{\rm F}}
\approx \alpha_{\rm eff} \, v(\Omega, N, R)$,  we calculate from
the time delay between the collapse of the cf peaks in the top and
bottom coils the time $t_{\rm b}$ and location $z_{\rm b}$ of the
turbulent burst. In the example of
Fig.~\ref{WallMediatedInstability} the measured delay of 30\,s
places the burst at a height $z_{\rm b} = 76\,$mm above the
orifice at time $t_{\rm b} = 32\,$s (measured from the moment when
the rotation drive reached $\Omega_{\rm f}$).

The analysis of the measured turbulent bursts allows us to
conclude that multiple bursts, which would occur almost
simultaneously, but in different locations along the rotating
column, are not frequent in the onset regime and clear candidates
of such events have not been identified. This conclusion is based
on the continuous well-behaved behavior of the measured data on
$V_{{\rm F}}$, $t_{\rm b}$, and $z_{\rm b}$. We believe that in
the onset regime the probability of the turbulent burst is still
small and the propagation of the vortex front so rapid that it is
unlikely for bursts to start at two different locations in
sufficiently close proximity in time.

\begin{figure}[t]
\begin{center}
\centerline{\includegraphics[width=0.5\linewidth]{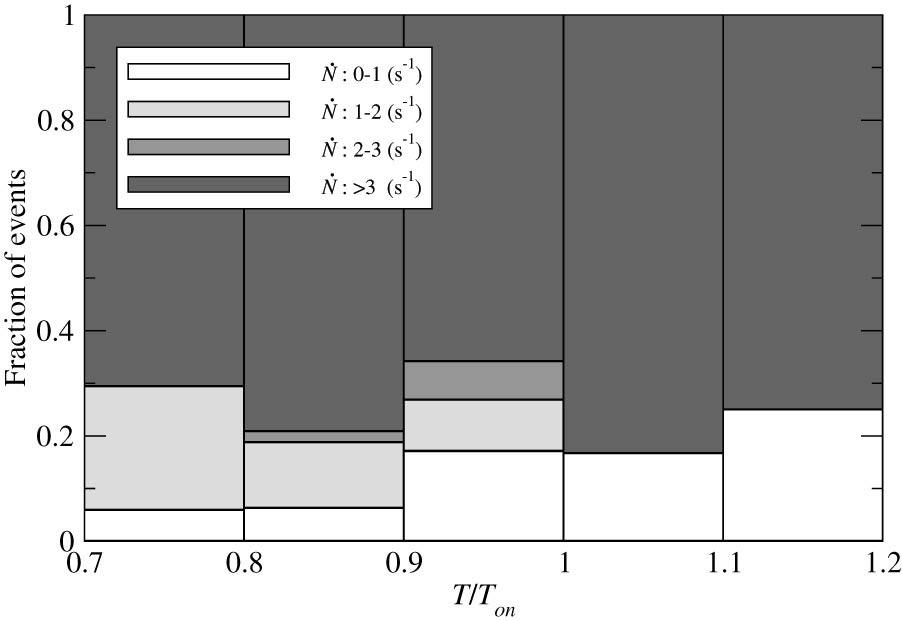}}
\caption{(Color online) $\dot{N}$ data from the onset regime in
Fig.~\protect\ref{PrecursorSlope}, arranged independently of the
$\Omega_{\rm i}$ value in four bins [$(0 < \dot{N} \leq
1\,$s$^{-1}$), (1,2), (2,3) and, ($\dot{N}
> 3\,$s$^{-1}$)] in different intervals of the normalized
temperature $T/T_{\rm on}$. This histogram of 140 data points
characterizes the tail of the $\dot{N}$ distribution towards slow
rates as a function of temperature and illustrates the frequency
at which such processes occur around $T_{\rm on}$. Although slow
rates $(\dot{N} < 3\,$s$^{-1}$) are virtually nonexistent below
$0.6\,T_{\rm on}$, this data set does not display a clearly
increasing rate with decreasing temperature (after averaging over
$\Omega_{\rm i}$).  } \label{NdotHistogram}
\end{center}
\vspace{-8mm}
\end{figure}

The central characteristic of the single-vortex instability in
Fig.~\ref{WallMediatedInstability} is the slowly decreasing cf
peak height which measures the rate of vortex formation $\dot{N}$.
A sufficiently long period of slow peak height decay for this type
of measurement is observed only in the onset regime, $T \sim
T_{\rm on}$. At lower temperatures the burst time $t_{\rm b}$
becomes very short after any small rotation increase $\Delta
\Omega$ and our measurement too slow for resolving such events.
The instability is easier to monitor at lower pressures where
longer burst times $t_{\rm b}$ are observed \cite{Precursor}.  In
Fig.~\ref{WallMediatedInstability}, $T_{\rm on} = 0.54\,T_{\rm c}$
is defined as the average of a series of measurements on the
transition to turbulence, which fit a normal distribution with a
half width $\sigma_{\rm T} = 0.02\,T_{\rm c}$. Here and in all
later examples, $T_{\rm on}$ is measured in each case separately
for the appropriate conditions of that particular measuring
situation, as was done in Fig.~\ref{RemOnsetTempLongCyl}.

\begin{figure}[t]
\begin{center}
\centerline{\includegraphics[width=1\linewidth]{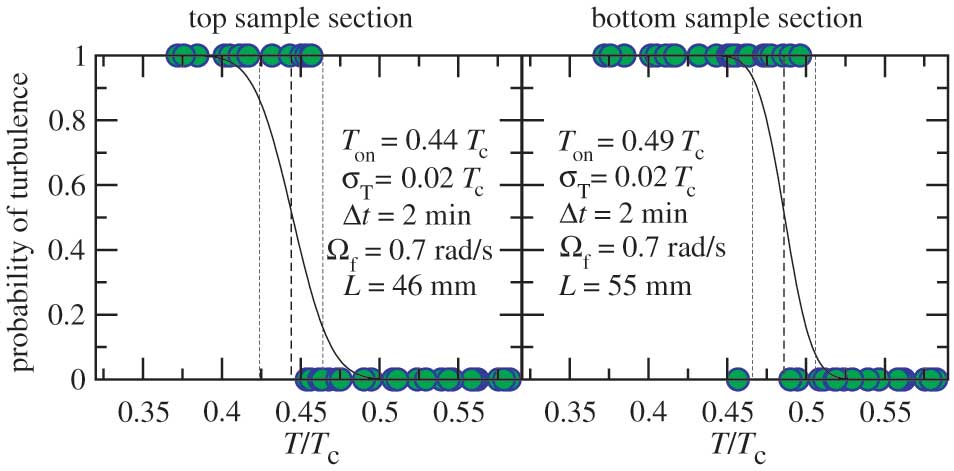}}
\caption{(Color online) Onset temperature of turbulence for the
upper and lower sections of the sample in Fig.~\ref{setup}. The
division into these two sections is described in
Ref.~\protect\cite{DynamicRemnants}. The measurements are
performed similar to those in
Fig.~\protect\ref{RemOnsetTempLongCyl}, starting from an initial
state at zero rotation with remanent vortices left over after an
annihilation period $\Delta t = 2\,$min. Comparing the two
sections, we recognize that $T_{\rm on}$ is higher in the bottom
section, {\it i.e.} the probability of turbulence at a given
temperature is enhanced in the bottom section. The main difference
between the two sections is the orifice in the lower part. This
suggests that the vicinity of the orifice is particularly
propitious for starting the single-vortex instability. The solid
curves are normal distributions with a half width $\sigma_{\rm T}
= 0.02\,T_{\rm c}$, centered around $T_{\rm on}$. }
\label{TopBottomComparisonTon}
\end{center}
\vspace{-8mm}
\end{figure}

To summarize, the collapse of the cf peak height in
Fig.~\ref{WallMediatedInstability} is caused by the arrival of the
vortex front. The front moves with a velocity which depends on the
number of vortices $N$ in the central cluster at height $z$,
before the front is about to pass at $z$. In an ideal case, where
the vortex instability occurs continuously and randomly in the
sample, one might expect that the site of the burst is randomly
distributed along the $z$ axis. Varying $\Omega_{\rm i}$, we can
change in a controlled manner the number of curved seed vortices
$\cal{N}_{\rm i}$ which connect initially to the cylindrical side
wall. In addition, by changing $\Delta \Omega$ or temperature, we
control respectively the applied flow velocity $v(\Omega,R,N)$ or
the damping $\alpha (T)$. By studying the dependence of $T_{\rm
on}$ on these variables one can analyze how they influence the
onset of turbulence (compare Figs.~\ref{RemOnsetTempLongCyl} and
\ref{TopBottomComparisonTon}). Next we are going to focus on the
properties of the precursor, represented by the approximately
linear $\dot{N} (t)$ in Fig.~\ref{WallMediatedInstability}, by
examining its characteristics in the onset regime $T \approx
T_{\rm on}$, such as the vortex formation rate $\dot{N}$, and the
distributions of the burst time $t_{\rm b}$ and the burst
locations $z_{\rm b}$.

\begin{figure}[t]
\begin{center}
\centerline{\includegraphics[width=0.8\linewidth]{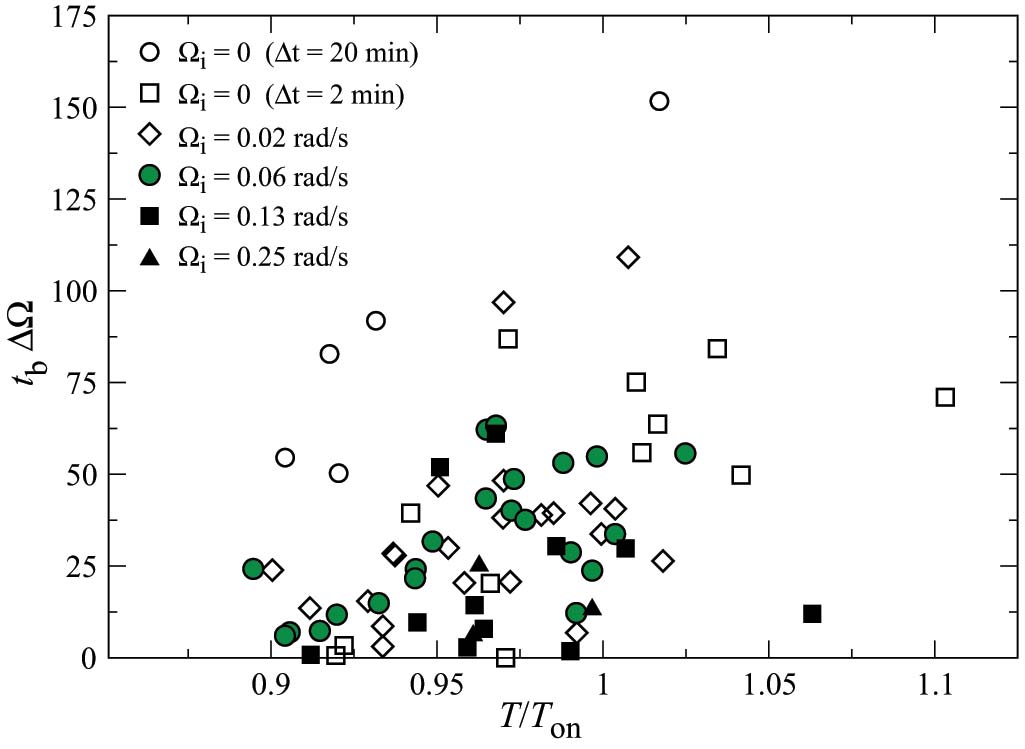}}
\caption{(Color online) Time $t_{\rm b}$ needed to start the
turbulent burst during continuous vortex generation at constant
rotation $\Omega_{\rm f} = \Omega_{\rm i} + \Delta \Omega$,
plotted as a function of temperature around the onset $T \approx
T_{\rm on}$. The data have been collected at five different values
of $\Omega_{\rm i}$ and three values of $\Delta \Omega = 0.7$,
0.9, and 1.3\,rad/s. Since $T_{\rm on}$ depends on both
$\Omega_{\rm i}$ and $\Delta \Omega$, in each case the appropriate
measured value of $T_{\rm on}$ is used for normalizing the
temperature axis. The data for $\Omega_{\rm i} = 0$ come from
measurements on remanent vortices with $\Delta \Omega =
0.7\,$rad/s, as in Fig.~\ref{RemOnsetTempLongCyl}. No A-phase
barrier field is applied in these measurements and thus the sample
is here twice as long as in Fig.~\ref{PrecursorSlope}. The plot
suggests that on average $t_{\rm b}$ increases with normalized
temperature $T/T_{\rm on}$ and decreases with $\Omega_{\rm i}$. }
\label{BurstTime}
\end{center}
\vspace{-8mm}
\end{figure}

\begin{figure}[b]
\begin{center}
\centerline{\includegraphics[width=0.327\linewidth]{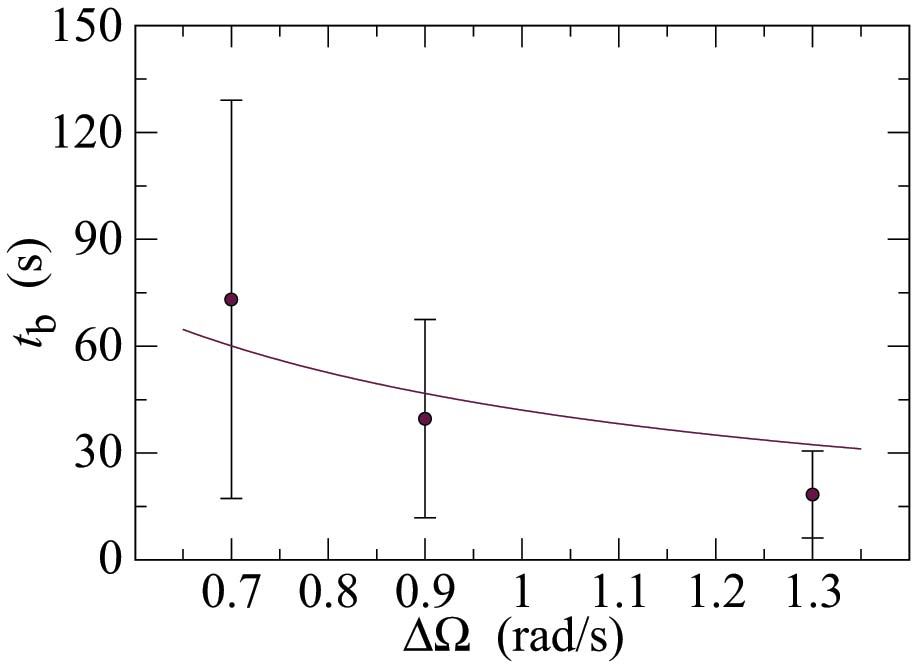}
\includegraphics[width=0.330\linewidth]{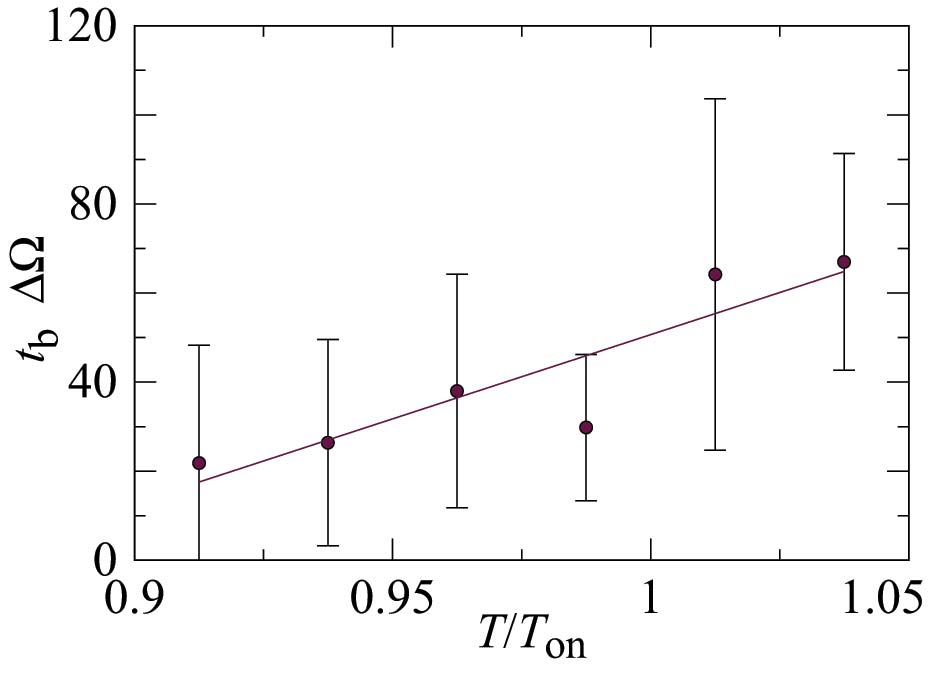}
\includegraphics[width=0.323\linewidth]{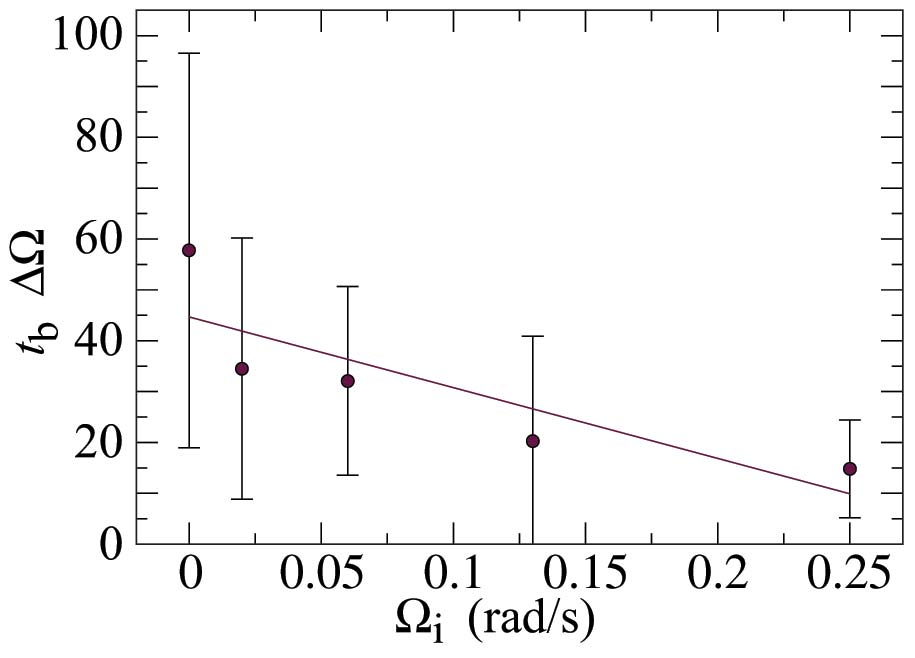}}
\caption{(Color online) Burst time $t_{\rm b}$, averaged for the
data in Fig.~\protect\ref{BurstTime} and analyzed for a dependence
on the applied flow velocity $\sim \Delta \Omega$ \textit{(left)},
on the normalized temperature $T / T_{\rm on}$ \textit{(center)},
and on the number of seed vortices $\sim \Omega_{\rm i}$
\textit{(right)}. The averaging has been done by dividing the data
points in bins as a function of the respective variable and by
denoting the standard deviation in each bin with uncertainty
limits. For instance, on the left we see that $t_{\rm b}$
decreases with increasing flow velocity $v(\Omega, N, R)$, since
this velocity is roughly expressed by the rotation increase
$\Delta \Omega$. Here the solid curve is the average $\langle
t_{\rm b} \rangle = 42/ \Delta \Omega$. }
\label{AveragedBurstTime}
\end{center}
\vspace{-8mm}
\end{figure}

{\bf Experimental results:}---Fig.~\ref{VorInstability} shows two
examples, after conversion from cf peak height to vortex number
$N(t)$. When vortex generation proceeds slowly, the rate $\dot{N}$
is initially of order $ 1 \,$vortex/s, but increases at later
times and becomes more nonlinear, until the vortex front passes
through the coil and $N$ jumps to nearly $N_{\rm eq}$. As seen in
Fig.~\ref{VorInstability}, generally the initial rate $\dot{N}
(t=0)$ increases with increasing $\Omega_{\rm i}$, since the
number of seed vortices $\cal{N}_{\rm i}$, which connect to the
cylindrical side wall in the initial equilibrium vortex state,
increases with $\Omega_{\rm i}$.

More statistics on the precursor are presented in
Fig.~\ref{PrecursorSlope}, collected from measurements similar to
those in Fig.~\ref{WallMediatedInstability}. The initial rate of
vortex generation $\dot{N}(t=0)$ is compared here in the onset
regime $T \sim T_{\rm on}$ for different $\Omega_{\rm i}$ and thus
for different number of evolving vortices $\cal{N}_{\rm i}$,
keeping the rotation drive $\sim \Omega_{\rm f} - \Omega_{\rm i} =
\Delta \Omega = 0.7\,$rad/s constant. In the inset the average of
the measurements is seen to depend roughly linearly on
$\Omega_{\rm i}$, as was concluded in the context of
Fig.~\ref{VorInstability}. The data in Fig.~\ref{PrecursorSlope}
includes only events with long burst times $t_{\rm b} \gtrsim
20\,$s, where the slope of the cf peak height with time can be
clearly determined (while events with $t_{\rm b} < 20\,$s can be
counted, but the value of $t_{\rm b}$ is not resolved). The data
is shown again by the histogram of Fig.~\ref{NdotHistogram}, but
now including also the fast events, where $t_{\rm b} < 20\,$s and
$\dot{N} > 3\,$vortices/s. Fig.~\ref{NdotHistogram} shows that, on
average, $\dot{N}$ increases in this data set with decreasing
temperature. In addition we observe no events with a slow
measurable $\dot{N}$ below $0.6 \leq T/T_{\rm on}$. We thus have
to conclude that extended precursory vortex generation with a
burst time $t_{\rm b} \gtrsim 20\,$s can only be observed in the
onset temperature regime $T \sim T_{\rm on}$.

\begin{figure}[t]
\begin{center}
\centerline{\includegraphics[width=0.7\linewidth]{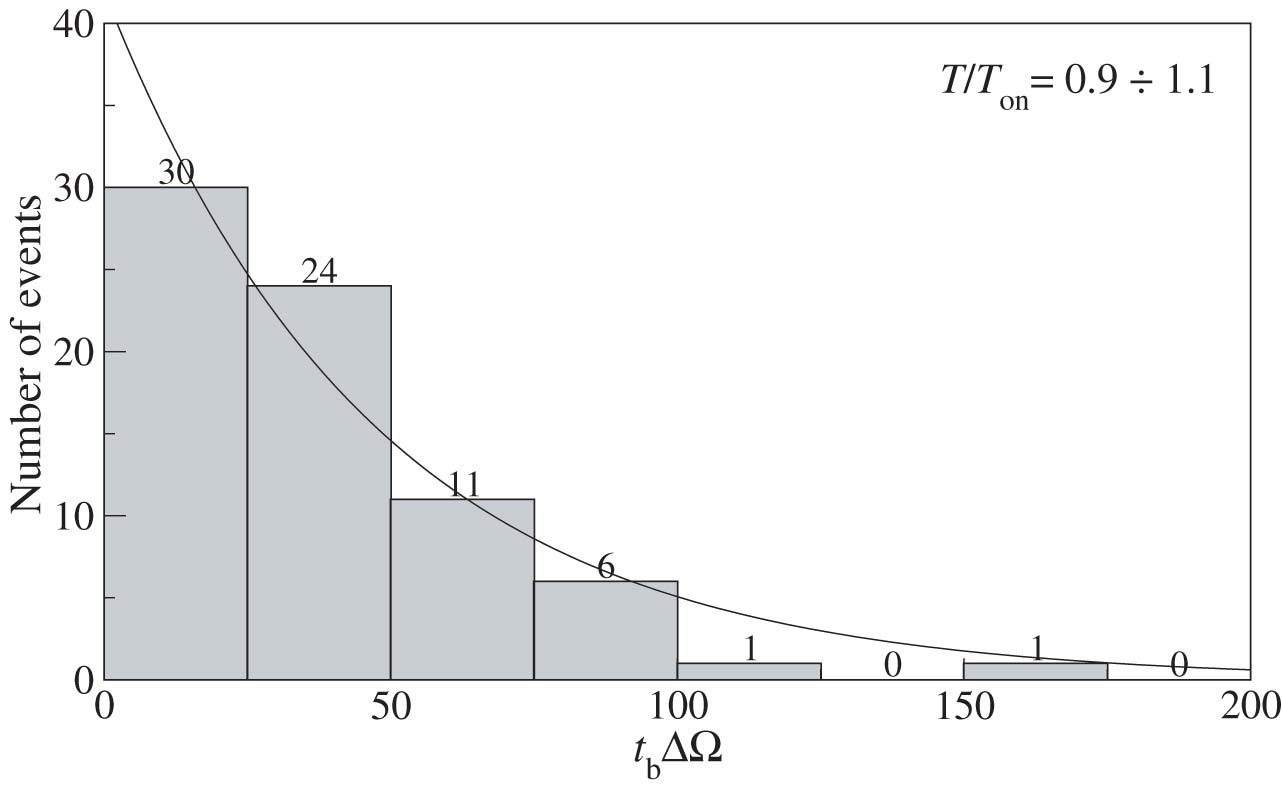}}
\caption{(Color online) Temporal distribution of turbulent bursts
in the onset temperature regime. The burst time data from
Fig.~\protect\ref{BurstTime} is here organized as a histogram for
events where $t_{\rm b} > 20\,$s is long enough to be resolved.
The solid curve is an approximation to the tail of the probability
distribution for $t_{\rm b}$ and represents the fitted exponential
$\propto \exp{(-t_{\rm b}\, \Delta \Omega/47)} \approx
\exp{(-t_{\rm b} / \langle t_{\rm b} \rangle}$, where $\langle
t_{\rm b} \rangle$ is the average $t_{\rm b}$ at given $\Delta
\Omega$ (Fig.~\ref{AveragedBurstTime}, \textit{left}). }
\label{BurstTimeTail}
\end{center}
\vspace{-6mm}
\end{figure}

In Fig.~\ref{BurstTime} a different data set with measurements on
the burst time $t_{\rm b}$ is examined. Since our measurement
captures efficiently only events with long burst times,
Fig.~\ref{BurstTime} represents the tail ($t_{\rm b} \geq 20\,$s)
of the burst-time distribution. These 73 data points are almost
half of all the measured turbulent events in the temperature
interval $0.9 < T/T_{\rm on} < 1.1$ in the present data set. Thus
events with $t_{\rm b} \gtrsim 20\,$s and prominent slow vortex
formation are relatively frequent close to $T_{\rm on}$. The
equations of vortex motion roughly scale with the effective
rotation drive, which in this case is $\sim \Delta \Omega$. As the
data have been collected with three different values of $\Delta
\Omega$, we plot the result as $t_{\rm b} \; \Delta \Omega$ which
corrects for differences in the drive sufficiently well. The main
conclusions to be drawn from this data are illustrated in the
three plots of Fig.~\ref{AveragedBurstTime}, where the averages
are shown as a function of the rotation increase $\Delta \Omega$,
normalized temperature $T / T_{\rm on}$, and the rotation velocity
of the initial equilibrium vortex state $\Omega_{\rm i}$. The
interpretation of these plots is that the burst time $t_{\rm b}$
decreases rapidly with increasing applied flow velocity $\sim
\Delta \Omega$, increases with temperature $T / T_{\rm on}$, and
decreases with with $\Omega_{\rm i}$, \textit{i.e.} the number of
seed vortices $\cal{N}_{\rm i}$. In Fig.~\ref{BurstTimeTail} the
tail of the burst time distribution is shown as a histogram,
indiscriminately for all $t_{\rm b}$ data in Fig.~\ref{BurstTime}.
As seen here, in the onset regime, $T \approx T_{\rm on}$, the
probability for large burst times decreases approximately
exponentially.

To summarize we conclude that both data sets, displayed in
Figs.~\ref{PrecursorSlope} and \ref{BurstTime}, illustrate that
well-resolved single-vortex instability events are distinguished
by slow vortex formation $\dot{N}(t=0)$ and a long burst time
$t_{\rm b}$. Such events can be found (i) in the onset temperature
regime, $T \approx T_{\rm on}$, (ii) by starting from a state with
a small number of seed vortices $\cal{N}_{\rm i}$, and (iii) by
keeping the applied flow velocity ($\sim \Delta \Omega$) as small
as possible.

\begin{figure}[t]
\begin{center}
\centerline{\includegraphics[width=0.7\linewidth]{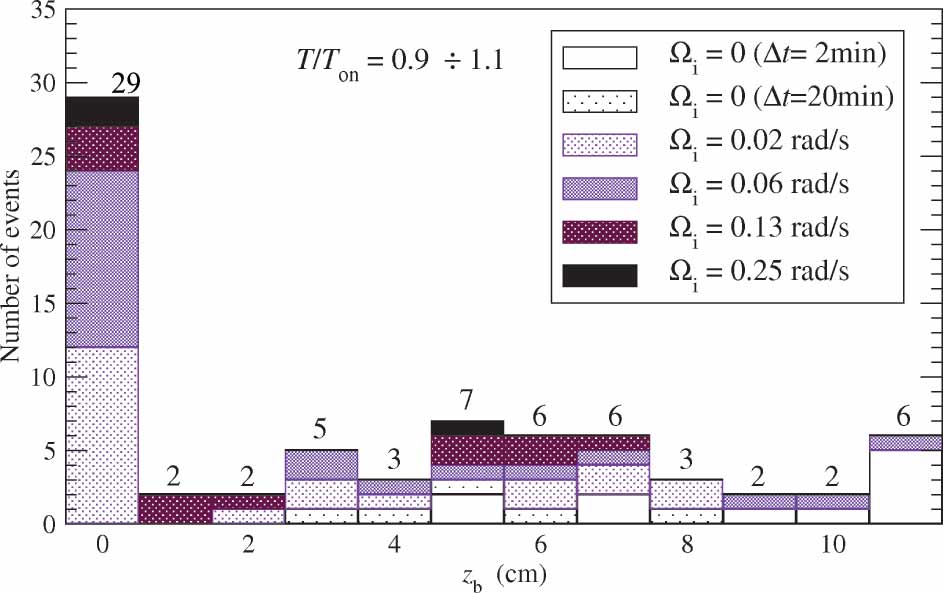}}
\caption{(Color online) Spatial distribution $z_{\rm b}$ of
turbulent bursts for the data in Fig.~\protect\ref{BurstTime},
organized as a histogram along the $z$ axis of the long sample in
Fig.~\protect\ref{setup}. The bursts appear to be randomly
distributed, except for a strong preference for the region below
the bottom coil. This additional mechanism, which enhances the
probability of the turbulent burst, we assume to be associated
with the presence of the orifice below the bottom coil. }
\label{BurstLocation}
\end{center}
\vspace{-8mm}
\end{figure}

In Fig.~\ref{BurstLocation} the spatial distribution of the
turbulent bursts from Fig.~\ref{BurstTime} is displayed along the
$z$ axis of the sample. As expected, the location $z_{\rm b}$ of
the burst is approximately evenly distributed along the column.
This supports the notion that the generation of new vortices
occurs randomly with equal probability along the entire cylinder.
The exception is a clear preference for the region below the
bottom coil. The breakdown of these events with $z_{\rm b} <
10\,$mm according to their $\Omega_{\rm i}$ values shows that the
orifice is a large perturbation for small vortex clusters. If
$\Omega_{\rm i} = 0.02\,$rad/s, then the cluster radius $R_{\rm o}
\approx R (\Omega_{\rm i}/\Omega_{\rm f})^{1 \over 2}$ is
approximately equal to the radius of the orifice. Even with
$\Omega_{\rm i} = 0.06\,$rad/s the two might be comparable, since
this comparison is affected by the centering of the orifice on the
bottom plate and the inclination of the cylinder and rotation
axes. At higher $\Omega_{\rm i}$ values the cluster apparently
covers the orifice more efficiently and its perturbing effect
fades away. Surprisingly, no cases of turbulent bursts are present
in Fig.~\ref{BurstLocation} which would have been started by
remanent vortices at the orifice (\textit{i.e.} with $\Omega_{\rm
i} = 0$), but this observation might change with a larger sample
of measured events.

\textbf{Discussion:}---We thus find that the presence of the
orifice promotes the probability of the single vortex instability
and concentrates turbulent bursts in the section below the bottom
detector coil. This is also seen in direct measurements on the
individual transition temperatures $T_{\rm on}$ of the top and
bottom sections of the long sample cylinder. In
Fig.~\ref{TopBottomComparisonTon} the cylinder has been divided in
two disconnected halves with a narrow transverse layer of
$^3$He-A, as first discussed in Ref.~\cite{KH-Instability}. The AB
interface acts as a barrier \cite{ROP} for vortices  and thus the
two sections can be sampled separately with the detector coils at
each end of the cylinder. Comparing the four $T_{\rm on}$
distributions for remanent vortices in
Figs.~\ref{RemOnsetTempLongCyl} and \ref{TopBottomComparisonTon},
it is seen that the $T_{\rm on}$ values are all distinctly
different, but in increasing order form a sequence which can now
be explained based on the previous conclusions. These four $T_{\rm
on}$ distributions have been measured in the same set of
measurements, without ever warming above dilution refrigerator
temperatures. This means that they are closely comparable and
representative of the same experimental environment. Nevertheless,
comparing the two distributions in
Fig.~\ref{TopBottomComparisonTon}, it is evident that a relatively
large difference of $0.05\,T_{\rm c}$ separates the $T_{\rm on}$
values of the top and bottom sections. This difference is large
enough so that the two distributions do not overlap. We conclude
that the probability of the turbulent burst increases at a given
temperature when the bottom section is included in the sample ---
or that the vicinity of the orifice is a particularly efficient
environment for starting the turbulent burst.

In earlier work \cite{OrificeFlow} it was proposed that a large
number of vortices might leak simultaneously through the orifice
as a sudden burst into the sample volume and that this is the
origin for the abrupt transition from the vortex-free to the
equilibrium vortex state. Below the orifice vortices are typically
present in the equilibrium vortex state in most situations, since
two quartz tuning fork oscillators are located there \cite{Fork}.
These provide ample opportunity for vortex pinning and trapping.
Moreover the bottom surface in this volume is a rough sintered
heat exchanger. However, considering
Figs.~\ref{RemOnsetTempLongCyl} and \ref{TopBottomComparisonTon}
together, a massive sudden leakage of vortices through the orifice
appears now questionable. Comparing the bottom section in
Fig.~\ref{TopBottomComparisonTon} to the corresponding plot for
the long sample in Fig.~\ref{RemOnsetTempLongCyl} (with $\Delta t
= 2\,$min on the right), we see that $T_{\rm on}$ is not
determined by only the orifice, but also depends on the length of
the cylinder above the orifice: The twice larger sample length $L$
of the long sample causes the onset $T_{\rm on}$ to increase by
$0.04\,T_{\rm c}$. This feature is consistent with the general
notion that in the long sample there are twice as many seed
vortices to begin with, the evolving vortices spend more time
spiralling along the cylinder wall, and are more likely to suffer
the instability, than in the short sample. In contrast, an event
with lots of vortices leaking through the orifice should be
insensitive to the length of the sample above the orifice and
seems less likely to explain the measurements on $T_{\rm on}$.

Nevertheless, Fig.~\ref{BurstLocation} shows that a large fraction
of the turbulent bursts occur in the vicinity of the orifice and
all measurements on the bottom section show a higher value of
$T_{\rm on}$ by about $0.04\,T_{\rm c}$ than an equivalent
measurement on the top section. This difference is visible also in
Fig.~\ref{TopBottomComparisonTon}. Thus we have to conclude that
the single-vortex instability is more likely at a given
temperature when the orifice is included in the sample and that
the bottom section therefore has a higher onset temperature than
the top. It is not clear at this point how the presence of the
orifice enhances the probability of the single-vortex instability,
since it affects both the initial configuration of seed vortices
and their later dynamics after the rotation increase. Here both
geometry as well as surface roughness could be important.

In contrast the lower value of $T_{\rm on}$ for the top section
indicates that an isolated cylinder (which in this case is closed
off by the AB interface barrier) displays a reduced probability
for the single-vortex instability to occur, in other words the top
section, with no obvious defects, is closer to an ideal
cylindrical sample. A further measurement on only the top section,
with a reduced number of initial seed vortices, obtained by
increasing the annihilation period $\Delta t$ from 2\,min to
20\,min, reduces $T_{\rm on}$ from $0.44\,T_{\rm c}$
(Fig.~\ref{TopBottomComparisonTon} left) to $0.39\,T_{\rm c}$.
This result is what we expect, based on the examples presented
above: $T_{\rm on}$ decreases if the number of seed vortices is
reduced.

To summarize, in the onset temperature regime ($T \approx T_{\rm
on}$) the single-vortex instability progresses sufficiently slowly
in about one third of the measured events so that it can be
recorded with our measurement techniques. It functions as the
precursor mechanism which generates new dynamically evolving
vortices, until a localized turbulent burst between interacting
vortices becomes possible in a short section (of length $\sim R$)
of the rotating column. The instability depends foremost on
temperature via the dynamic parameter $\zeta = (1-\alpha^{\prime})
/ \alpha$, which for superfluids is the equivalent of the Reynolds
number of viscous hydrodynamics, namely the ratio of the inertial
and dissipative forces \cite{ROP}. Our measurements in the onset
regime can be interpreted in terms of the probability of a single
dynamically evolving vortex to undergo the instability and to
create a new vortex loop which in turn starts to evolve. At a
given temperature the probability to achieve bulk turbulence
depends on the applied cf velocity at the container boundary, on
the length of the trajectory over which the vortex end travels
along the boundary, and on the total number of vortices which
simultaneously are dynamically evolving.

An earlier explanation of the onset of superfluid turbulence was
provided by Klaus Schwarz who in 1993 concluded (based on his own
work and that of others) that a set of several vortex mills is
required to start and maintain turbulence in channel flow
\cite{Schwarz}. These vortex mills need to act in parallel and
have to be located close to the entrance of the flow channel. Our
results now show that vortex mills are not necessary to start
turbulence and that there exists a more fundamental instability
mechanism, namely the single-vortex instability. In principle, the
characterization of this instability in
Figs.~\ref{RemOnsetTempLongCyl}, and \ref{WallMediatedInstability}
-- \ref{TopBottomComparisonTon} can be compared to simulation
calculations, to reconstruct a more detailed understanding. A step
towards this goal is taken in the next section, where the
instability mechanism is studied in numerical calculations.

\section{NUMERICAL CALCULATIONS ON SINGLE-VORTEX INSTABILITY}
\label{VortexFilamentCalculation}

{\bf Numerical method:} Our calculations \cite{Simulation} are
carried out using the vortex filament model introduced by Schwarz
\cite{Schwarz88}. With Biot-Savart integration along all vortex
lines the superfluid velocity field from vortices is obtained from
\begin{equation}
\mathbf{v}_{{\rm s},\omega}({\bf r},t) = \frac{\kappa}{4\pi} \int
\frac{(\mathbf{s-r}) \times d\mathbf{s}}{|\mathbf{s-r}|^3}.
\label{Biot-Savart}
\end{equation}
The line integral is taken along the vortices and ${\bf s}(\xi,t)$
denotes the location of the vortex core at time $t$, while $\xi$
is measured along the arc length of the vortex core. In the
presence of solid boundaries the total superfluid velocity field,
$\mathbf{v}_\mathrm{s} = \mathbf{v}_{\mathrm{s},
\omega}+\mathbf{v}_\mathrm{b}$, is modified by the boundary
induced velocity $\mathbf{v}_\mathrm{b}$. At a plane boundary one
can use image vortices to satisfy the requirement of zero flow
through the boundary, $\hat{\bm{n}} \cdot \mathbf{v}_\mathrm{s} =
0$, where $\hat{\bm{n}}$ is the unit vector along the surface
normal. More generally we obtain $\mathbf{v}_\mathrm{b} =
\nabla\Phi$ by solving the Laplace equation $\nabla^2\Phi = 0$
combined with the requirement that at the boundary $\hat{\bm{n}}
\cdot \nabla \Phi = - \hat{\bm{n}} \cdot \mathbf{v}_{\mathrm{s},
\omega}$. No surface pinning or surface friction is included, the
boundaries are assumed ideal which, as far as known, is not in
contradiction with measurements. Mutual friction in the bulk
superfluid is included using the equation of motion for the vortex
element at ${\bf s}(\xi,t)$
\begin{equation}
{\bf v}_{\rm L} = \frac{d\mathbf{s}}{dt} =\mathbf{v}_{\rm s}
+\alpha \mathbf{s}' \times (\mathbf{v}_{\rm n}-\mathbf{v}_{\rm s})
-\alpha' \mathbf{s}' \times [\mathbf{s}'\times(\mathbf{v}_{\rm
n}-\mathbf{v}_{\rm s}
)]\,. \label{vl}
\end{equation}
where the vector ${\bf s}'=d{\bf s}/d\xi$ is the local tangent to
the vortex at the point ${\bf s}(\xi,t)$. For the mutual friction
parameters $\alpha (T,P)$ and $\alpha'(T,P)$ we use the 29\,bar
data measured in Ref.~\cite{bevan}.

In practice, the implementation of the Biot-Savart integration is
performed adaptively, {\it i.e.} the number of discretization
points along a vortex in evaluating the Biot-Savart integral is
increased recurrently, until the required accuracy is obtained.
The vortex is split along its core into line segments whose length
is adjusted such that shorter segments are used in places where
the vortex is more curved or the counterflow is large (enabling
smaller wavelength Kelvin-waves). The smallest segment length
$\Delta\xi$ limits the time step $\Delta t$ which is used to solve
the time development of the tangle with the classical 4th order
Runge-Kutta method. The solution of the Laplace equation is
obtained by discretizing the potential $\Phi$ within the cylinder
(typical grid size, \textit{eg.} radially, $\Delta r = R/15$). The
resulting sparse matrix equation is then solved at each time step,
while the spatial derivatives are approximated with finite
differences. This means that the continuity equation for the
superfluid velocity is not accurately satisfied. Nevertheless,
this scheme is an improvement over much of the earlier work.

To solve for ${\bf v}_{\rm b}$, one needs to make sure that
vortices meet the boundaries perpendicularly and that in
Eq.~(\ref{Biot-Savart}) one integrates along vortices which form
closed loops, as noted by Schwarz \cite{Schwarz85}. The latter
requirement is implemented by extending the vortices, which
terminate perpendicularly on the boundary, to infinity with
straight vortex line sections. A vortex reconnection is performed
when two vortices approach each other closer than the maximum
resolution (=$\Delta\xi \sim R/50$ typically, measured along the
vortex core), provided that the resulting configuration has
reduced length and represents thus a lower energy state. Generally
the maximum resolution has minor effect on the results in the
intermediate temperature regime. Increased resolution slows down
the calculations and results in larger numbers of tiny vortex
loops which in any case rapidly disappear owing to the finite
mutual friction damping. Nevertheless, a sufficiently fine
resolution is needed to display Kelvin-wave excitations, and
towards low temperatures the resolution needs to increase rapidly.
In solving the Laplace equation for the boundary condition, a
coarser resolution must be tolerated, to avoid too large memory
consumption.

\begin{figure}[t]
\begin{center}
\centerline{\includegraphics[width=0.8\linewidth]{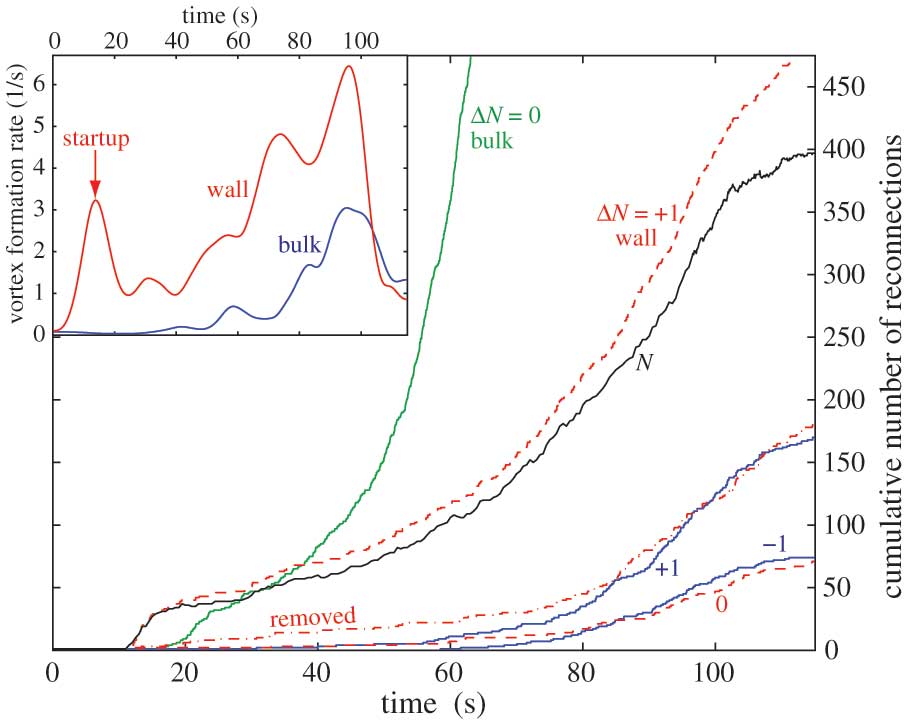}}
\caption{(Color online) {\it (Main panel)} Calculated build up in
cumulative number of vortices and their reconnections in a
rotating cylinder: (0, solid curve) reconnections in the bulk
which do not change $N$, (+1, dashed) surface reconnections which
add one new vortex loop, (N) total number of vortices, (removed,
dash-dotted) small loops which form in reconnections mainly close
to the cylindrical wall, but which are contracting and are
therefore removed, (+1, solid) bulk reconnections which add one
vortex, (-1, solid) bulk reconnections which remove a vortex, (0,
dashed) reconnections at the boundary which do not change $N$.
{\it (Insert)} Averaged number of reconnections per second on the
cylindrical boundary and in the bulk which add one vortex. The
large initial peak in the boundary rate represents the starting
burst which is required to start vortex formation. Parameters: $R
= 3\,$mm, $L = 10\,$mm, $\Omega = 0.9\,$rad/s, and $T=
0.35\,T_{\rm c}$ (where $\alpha = 0.095$ and $\alpha^{\prime} =
0.082$ \cite{bevan}).} \label{VorReconRotCyl}
\end{center}
\vspace{-8mm}
\end{figure}

Fig.~\ref{RemEvolution} is an example of how calculations can be
used to illustrate and interpret measurements. This calculation is
performed at $0.38\,T_{\rm c}$ and conserves the number of
vortices during their evolution, after increasing the rotation
velocity from zero to $\Omega_{\rm f}$. Comparing to
Figs.~\ref{RemOnsetTempLongCyl} and \ref{TopBottomComparisonTon},
we note that in most experimental cases $0.4\,T_{\rm c}$ is below
$T_{\rm on}$, so that the single-vortex instability would
interfere and would lead to turbulence. The origin of this
difference between calculations and experiments has not been
resolved. To start vortex generation in the calculation, often a
specially designed unstable starting configuration is required,
which creates a larger number of interacting vortices
\cite{Precursor}.

{\bf Vortex generation in rotating flow:}---In
Fig.~\ref{VorReconRotCyl} the generation of vortices is studied in
a short circular cylinder with diameter comparable to height.
Account is kept of all reconnection processes as a function of
time while the sample is evolving towards its final state. Vortex
formation is initially started from a single vortex ring which is
placed in the plane perpendicular to the rotation axis at height
$0.2\,L$ slightly off center, to break cylindrical symmetry (see
Ref.~\cite{Precursor}). This is an unstable configuration where
Kelvin waves of large amplitude immediately form and reconnect at
the cylindrical wall. The end result is a sudden formation of
roughly 30 vortices which have one end on the bottom end plate and
the other moving in spiral trajectory along the cylindrical wall.
After this initial burst the spontaneous evolution is followed in
the calculation, the formation of new vortices is noted, and the
reconnections of different type are classified.

\begin{figure}[t]
\begin{center}
\centerline{\includegraphics[width=0.7\linewidth]{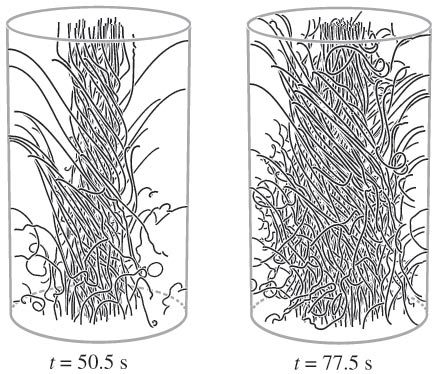}}
\caption{Two snapshots from the calculation in
Fig.~\protect\ref{VorReconRotCyl}. In both cases helically twisted
younger vortices can be seen on the outer circumference of the
cluster and older straighter vortices in the center. Primarily
surface reconnections contribute to the formation of new vortices
at $t < 100\,$s which explains the many short loops outside both
clusters. } \label{VorConfigRotCyl}
\end{center}
\vspace{-8mm}
\end{figure}

In Fig.~\ref{VorConfigRotCyl} we see snapshots of vortex
configurations after 50\,s and 80\,s. Recently formed vortices are
here on the outer circumference in helical configuration, while
further inside the cluster the vortices are gradually relaxing
towards rectilinear lines. Outside the cluster closer to the
cylindrical wall one can see loops of Kelvin waves, small
separated loops with both ends of the vortex on the cylindrical
wall, and even closed vortex rings (lower right corner at
$t=50\,$s).

Returning to Fig.~\ref{VorReconRotCyl}, we note that after the
initial burst of the first $\sim 30$ vortices $N$ increases first
gradually, but after about 50\,s the rate $\dot{N}$ picks up.
During the first 50\,s reconnections in the bulk do not contribute
to the generation of new vortices, but later such processes also
start appearing. Surprisingly however, even during the later phase
a reconnection of a single vortex at the cylindrical wall, while a
Kelvin wave expands along this vortex, remains the dominant
mechanism of vortex generation. This is seen from the fact that
the curve for $N$ follows closely that of the successful surface
reconnections (dashed curve marked as ``+1"). In comparison such
surface reconnections are few in which a small loop is created,
but which later shrinks away, for instance, because it is
adversely oriented with respect to the azimuthal cf (dash-dotted
``removed" curve [which actually includes such small loops from
both the surface and the bulk]). The dashed curve denoted as ``0"
refers to processes where a closed vortex ring from the bulk
drifts against the cylindrical wall. Such cases do not change the
value of $N$. They require successful vortex-generating
reconnections in the bulk and consequently the dashed ``0" curve
emerges only after the solid ``+1" curve has acquired sufficient
slope.  In contrast the solid ``0" curve represents reconnections
in the bulk between two different vortices which after the first
40\,s rapidly becomes the most frequent event. These inter-vortex
reconnections do not lead to changes in $N$ and are primarily
associated with processes occurring between the twisted vortices
within the bundle, where they help to increase the polarization of
the vortices along the rotation axis. One might ask whether such
bulk reconnections nevertheless emit Kelvin wave excitations which
then propagate to the boundary and lead to loop formation and
reconnections there. At present there is no clear evidence of
that.

\begin{figure}[t]
\begin{center}
\centerline{\includegraphics[width=0.5\linewidth]{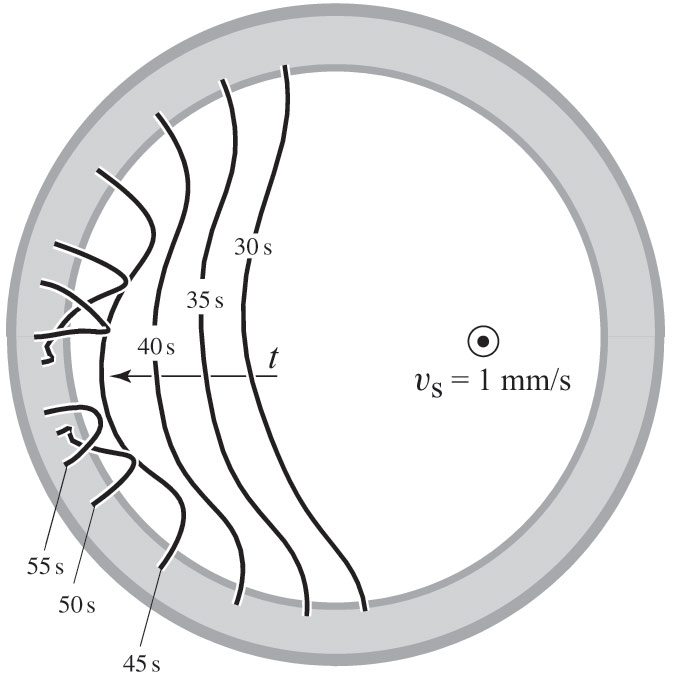}}
\caption{(Color online) Boundary-mediated vortex formation in pipe
flow. The calculation shows how a seed vortex travels across the
cross section of the pipe after its release. The perspective view
looks straight into the pipe against the flow, the inner ring is
at $z= 20\,$mm and the outer at $z= 70\,$mm. Originally at $t=0$
the center of the seed vortex was at $z=0$  closer to the pipe
wall on the right. The vortex drifts both along the pipe (towards
the viewer) and across the flow channel (from right to left). Its
center section adopts the curvature of the pipe and annihilates
(at $45 < t < 50\,$s). The small loops at both ends survive as
independent vortices, they are formed to satisfy the boundary
condition. Reconnection kinks appear on both loops (at $t=50\,$s),
but these do not expand. The two loops reorient themselves with
respect to the flow and then drift across the flow channel in the
opposite direction. In this way the number of vortices has
increased by one.  The repolarization of the two new loops happens
within the time span 50 -- 55\,s, as seen in
Fig.~\ref{TubeFlowVorFlip}. The present figure shows the start of
the calculation in Fig.~\protect\ref{VorReconTubeFlow} and thus
the parameters are here, as well as in Fig.~\ref{TubeFlowVorFlip},
the same as in Fig.~\protect\ref{VorReconTubeFlow}. }
\label{TubeFlowVorForm}
\end{center}
\vspace{-8mm}
\end{figure}

The insert in Fig.~\ref{VorReconRotCyl} compares the rates of
vortex generation from reconnections at the wall and in the bulk.
The dominant role of wall reconnections is compelling. Other
similar calculations lead to the same conclusion: The reconnection
of a single vortex at the cylindrical wall is the most important
mechanism for the generation of new vortices. This process was
illustrated by means of a numerical example in
Ref.~\cite{Precursor}, while the task of Fig.~\ref{VorReconRotCyl}
is to provide quantitative estimates of the relative frequencies
of successful vortex-generating reconnections at the wall and in
the bulk.

The second important consideration is correspondence with
measurement. The obvious difference between calculation and
measurement is the ease with which new vortices are generated in
experiment below $T_{\rm on}$, whereas in
Fig.~\ref{VorReconRotCyl} the rate of vortex generation remains
modest. No clearly identifiable turbulent burst can be
distinguished in Fig.~\ref{VorReconRotCyl}. The same calculation
at a lower rotation velocity of 0.8\,rad/s gives a qualitatively
similar result, both with respect to $\dot{N}(t)$ and the break
down in different types of reconnections, except that all rates
are smaller. After about 150\,s both the surface and bulk rates
turn off simultaneously and vortex generation stops at $N \approx
290$ vortices, well below the equilibrium number $N_{\rm eq}
\approx 570$. Clearly in this example no turbulent burst takes
place, which would boost the vortex number up to $N_{\rm eq}$. In
Fig.~\ref{VorReconRotCyl} at 0.9\,rad/s the calculation has been
continued to 115\,s and $N \approx 400\,$vortices, where vortex
generation starts to slow down, again well short of $N_{\rm eq}
\approx 650\,$vortices. The calculations are time consuming which
limits our possibilities to obtain a more comprehensive
understanding of their predictions.

\begin{figure}[t]
\begin{center}
\centerline{\includegraphics[width=0.5\linewidth]{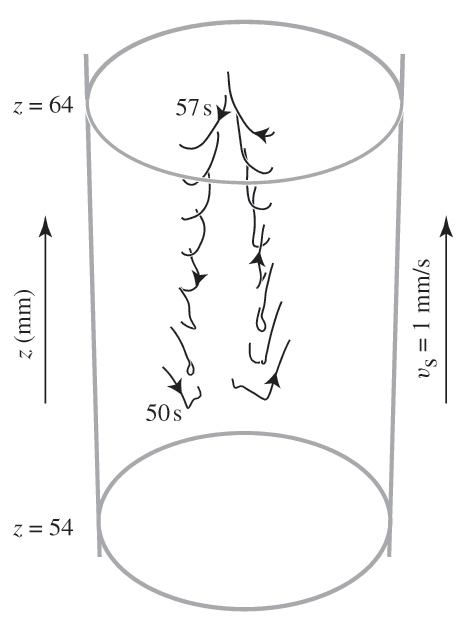}}
\caption{(Color online) Reorientation of the two new vortex loops
in Fig.~\ref{TubeFlowVorForm}, shown at 1\,s intervals. On the
bottom ($t= 50\,$s) the two loops have just formed on the back
wall of the pipe and have the wrong orientation with respect to
flow, while at the top ($t= 57\,$s) they have changed their
orientation and start expanding towards the viewer. The arrows on
the loops show the orientation of the circulation vector
$\mathbf{\kappa}$. The parameters are the same as in
Fig.~\protect\ref{VorReconTubeFlow}.} \label{TubeFlowVorFlip}
\end{center}
\vspace{-8mm}
\end{figure}

It thus appears as if some mechanism is missing from the
calculations in comparison to experiment, which makes vortices
more unstable and adds to the vortex generation rate. The
difference is less likely to reside in the bulk than on the
cylindrical wall, where the condition of an ideal solid boundary
should be examined closer. Nevertheless, at low vortex density
Kelvin-wave formation on a single vortex followed by a
reconnection at the surface is the only efficient mechanism for
generating a new vortex. The problem is complex, since Kelvin-wave
formation depends on the flow velocity and the orientation of the
flow with respect to the vortex, which in turn change continuously
while the vortex moves and the Kelvin wave itself starts to
propagate. It appears that with respect to the reconnection of
Kelvin-wave loops at the wall, the rotating cylinder with circular
cross section is a particularly stable flow geometry. The next
example calculates the equivalent of Fig.~\ref{VorReconRotCyl} for
linear flow in a circular pipe. Here vortices turn out to be less
stable than in rotating flow. The reason is the enhanced role of
reconnections at the wall.

{\bf Vortex generation in pipe flow:}---Technically a measurement
with seed-vortex injection in vortex-free linear pipe flow of
superfluid $^3$He-B is a demanding task; so far such measurements
have not been performed. Nevertheless, we present here
calculations on a circular straight tube which is initially
vortex-free. The calculations are performed similar to those above
on rotating flow, but by approximating the boundary conditions
with the faster image vortex techniques. The cf is enforced by
imposing on the superfluid component flow at constant velocity
over the cross section of the pipe.

\begin{figure}[t]
\begin{center}
\centerline{\includegraphics[width=0.8\linewidth]{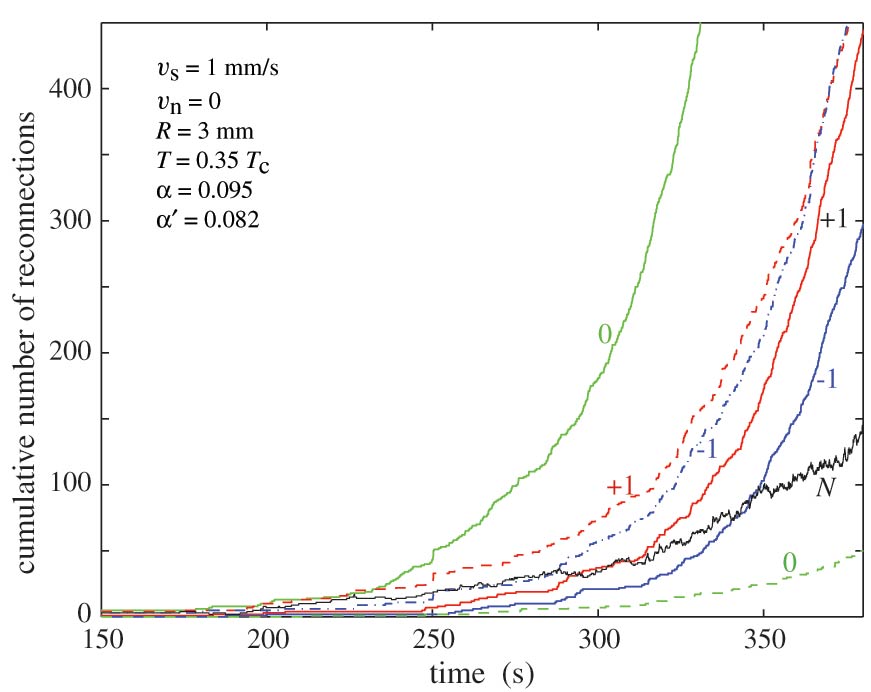}}
\caption{(Color online) Calculated build up in the cumulative
number of vortices and reconnections in linear pipe flow with
circular cross section: (0, solid curve) reconnections in the bulk
which do not change $N$, (+1, dashed) reconnections at the
boundary which add one new vortex loop, (-1, dash-dotted) very
small contracting loops both on the surface and in the bulk which
approach the resolution limit and are removed, (+1, solid)
reconnections in the bulk which add one vortex, (-1, solid)
reconnections in the bulk between a closed loop and another vortex
which remove one vortex, (N) total number of vortices, and (0,
dashed) reconnections at the boundary which do not change $N$.
} \label{VorReconTubeFlow}
\end{center}
\vspace{-8mm}
\end{figure}

Technically such measurements could be set up in the following
manner: Suppose that both the entrance and the exit of the tube
are covered with a superleak which prevents the flow of the normal
component. The superfluid component is forced into motion with a
piston acting on a large reservoir in front of the
superleak-covered entrance of the flow tube. Obviously in a real
experiment of this kind large numbers of vortices would be created
in the superleak. These would continuously flood the tube, as long
as the flow at constant mass rate is maintained. Such a
measurement would not be informative about vortex generation and
the onset of turbulence. To avoid this problem, the tube could be
bent to a closed ring, in which the flow is created by rotating,
in the same way as in the rotating cylinder.

In our numerical calculation we assume ideal laminar flow through
the circular straight tube. To start vortex generation, we place
one straight vortex line in the flow, which stretches from wall to
wall across the tube, slightly tilted from the perpendicular
plane, to break the symmetry. It turns out that the later
evolution of the seed vortex is rather insensitive to its initial
configuration and that our results do not depend materially on how
the seed vortex was originally placed in the tube. Experimentally
seed vortex injection can be achieved by creating vortex rings in
applied flow by means of the neutron capture reaction in neutron
irradiation \cite{NeutronInjection}.

\begin{figure}[t]
\begin{center}
\centerline{\includegraphics[width=0.7\linewidth]{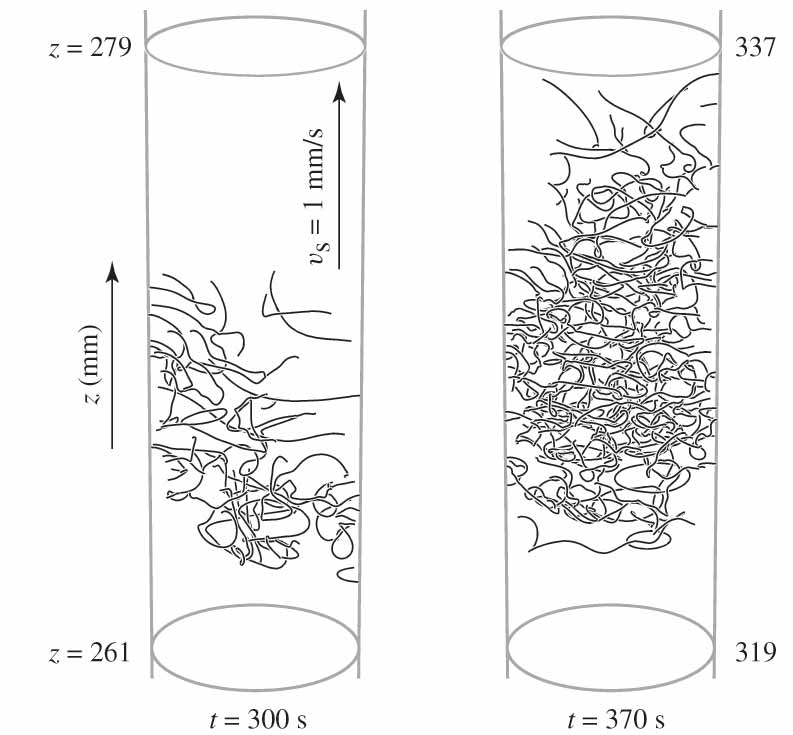}}
\caption{Two snapshots from the evolution of the vortex tangle in
pipe flow in Fig.~\protect\ref{VorReconTubeFlow}. {\it On the
left} the tangle is shown 300\,s after the start from one seed
vortex and {\it on the right} after 370\,s. During their evolution
the vortices have drifted in the pipe along the $z$-axis
(dimensions are given in mm), where $z=0$ was fixed at the middle
point of the seed vortex at $t=0$. The parameters are the same as
in Fig.~\protect\ref{VorReconTubeFlow}.} \label{VorConfigTubeFlow}
\end{center}
\vspace{-8mm}
\end{figure}

In Fig.~\ref{TubeFlowVorForm} we examine the trajectory of the
seed vortex along and across the flow channel.  A flat velocity
distribution $v_{\rm s} = 1\,$mm/s is imposed on the superfluid
component over the cross section (while the motion of the normal
component is clamped by the superleaks or by a large kinematic
viscosity, $v_{\rm n} = 0$). The vortex drifts with roughly the
velocity $v_{\rm s}$ along the pipe downstream (since
$\alpha^{\prime} \ll 1$), while it also moves transverse across
the tube, driven by the dissipative mutual friction force $\propto
\alpha v_{\rm s}$. The consecutive configurations of the vortex
are shown in Fig.~\ref{TubeFlowVorForm} at 5\,s intervals. Owing
to the boundary condition on the wall, the vortex bows out in the
center and mimics the curvature of the circular pipe wall, while
it traverses across the entire cross section. Ultimately its
center section, which is now aligned along the pipe wall,
annihilates. Only one small loop from both ends of the original
vortex remains, the vestiges from the boundary condition. In this
example both of these end loops manage to reorient themselves with
respect to the flow direction (Fig.~\ref{TubeFlowVorFlip}) and
then start an expanding motion in the opposite direction across
the flow. A successful reorientation is not always the case; often
one end loop may simply contract and annihilate. However, in
Fig.~\ref{TubeFlowVorFlip} the number of vortices starts to grow
continuously from one single seed vortex. Here the transverse
flight time across the flow is approximately 60\,s ($\sim 2
R/(\alpha v_{\rm s}$). Thus after the first 60\,s we have two
vortices, four vortices after $\sim 120\,$s, and after $\sim
180\,$s $N$ starts increasing more rapidly, as seen in
Fig.~\ref{VorReconTubeFlow}.

\begin{figure}[t]
\begin{center}
\centerline{\includegraphics[width=0.55\linewidth]{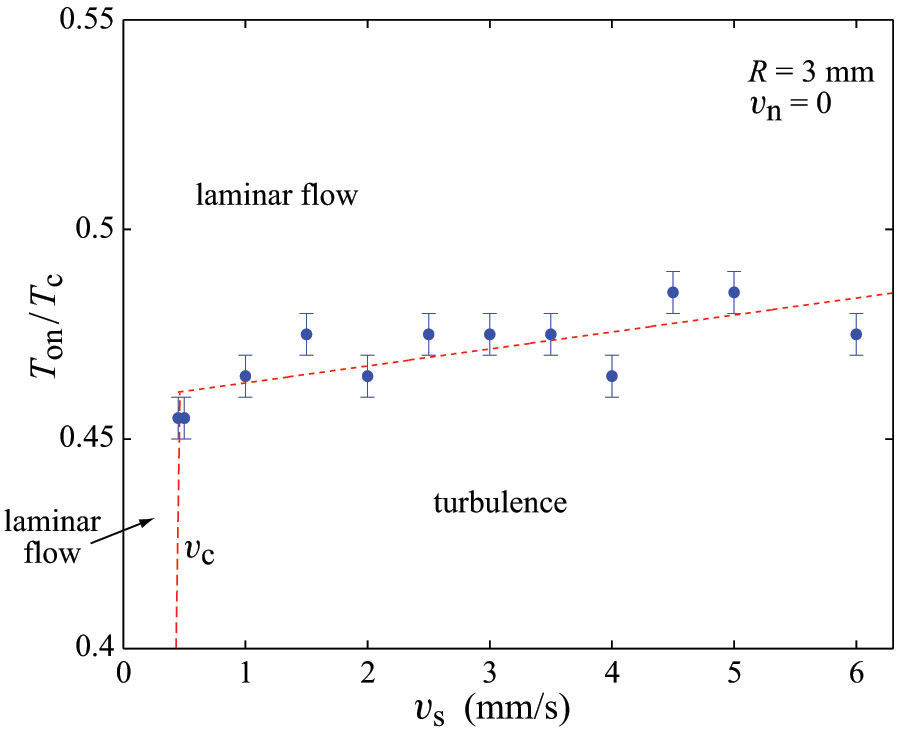}}
\caption{(Color online) Phase diagram of flow states and
dependence of the onset temperature of turbulence on flow velocity
in linear pipe flow. The calculations are started with a single
initial vortex, as in Fig.~\protect\ref{VorReconTubeFlow}. At low
velocities below the almost vertical critical boundary sustained
dynamic vortex generation cannot be maintained at any temperature
down to below $0.3\,T_{\rm c}$ and a state of turbulent flow is
not created. At higher velocities above this critical boundary
sustained turbulence exists at and below the onset temperature
plotted on the vertical scale. The parameters are the same as in
Fig.~\protect\ref{VorReconTubeFlow}.} \label{PipeOnsetTemp}
\end{center}
\vspace{-8mm}
\end{figure}

Comparing Figs.~\ref{VorReconRotCyl} and \ref{VorReconTubeFlow} it
is evident that for pipe flow -- in contrast to rotating flow --
there is no difficulty in starting turbulence from a single seed
vortex in the simulation calculation. The reason is the difference
in flow geometry: As seen in Fig.~\ref{TubeFlowVorForm}, for pipe
flow the generation of new vortices does not depend only on the
successful expansion of Kelvin-waves, but is aided by the boundary
condition which gives rise to characteristic end loops.
Nevertheless, in both calculations it is the interaction with the
wall which is responsible for the early phase of vortex
generation, in Fig.~\ref{VorReconTubeFlow} up to about 300\,s.
After about 360\,s the generation and annihilation of loops in
surface reconnections compensate each other and from here onwards
the generation in bulk becomes mainly responsible for the
production of new vortices.

Interestingly, also in Fig.~\ref{VorReconTubeFlow} reconnections
between two different vortices in the bulk, which do not directly
lead to new vortices, soon dominate over all other processes.
These bulk reconnections abound, when a sufficiently dense tangle
has formed and vortices traverse across the flow in both
directions, as seen in the two snapshots in
Fig.~\ref{VorConfigTubeFlow}. This leads to rapidly changing
configurations in the tangle. Furthermore, since all vortices
travel downstream, a turbulent plug is formed which hardly if at
all spreads upstream (pinning is excluded from our numerical
model). Ultimately, when the vortex plug reaches the superleak at
the exit of the flow tube, the vortices are annihilated and the
original state of vortex-free flow reappears.

It is instructive to note that turbulent plugs have been observed
in the flow of superfluid $^4$He along a straight capillary tube
with circular cross section \cite{VanBeelen}. In these
measurements temperature fluctuations are registered along the
tube in steady state flow conditions in some velocity range. The
fluctuations are interpreted to arise from the flow of turbulent
plugs along the tube. The plugs consist of a tangle of quantized
vortices, they extend over a limited length of the tube, and
display relatively sharp fronts with the laminar flow sections.
However, presumably in the case of $^4$He-II the transition from
laminar to turbulent flow is not governed by the supply of
vortices, as discussed here for pipe flow of $^3$He-B, but the
formation of the plugs is regulated by the associated fluctuations
in flow velocities, similar to the formation and decay of
turbulent plugs in viscous pipe flow \cite{Mullin} (where the
plugs are formed from classical eddies).

\begin{figure}[t]
\begin{center}
\centerline{\includegraphics[width=0.55\linewidth]{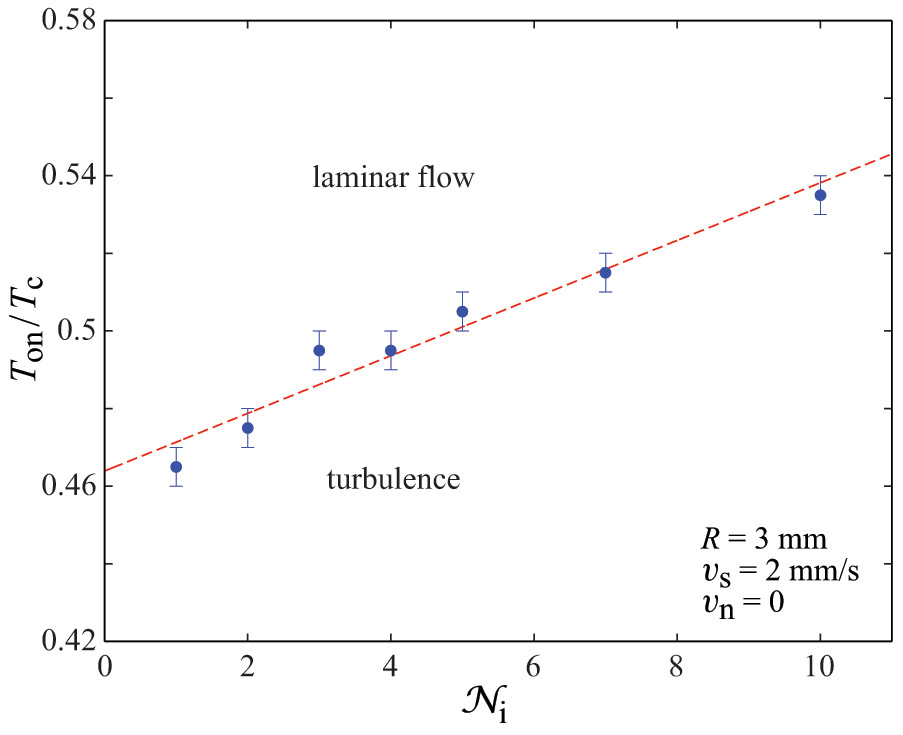}}
\caption{(Color online) Dependence of the onset temperature of
turbulence on the number of seed vortices in linear pipe flow. The
calculations are started with $\mathcal{N}_{\rm i}$ seed vortices
which are placed transverse across the pipe at equidistant
separation from each other, with their center points in the same
cross section of the tube at $z =0$. Other conditions are the same
as in Fig.~\protect\ref{VorReconTubeFlow}, except for $v_{\rm s} =
2\,$mm/s. At larger values of $\mathcal{N}_{\rm i}$ (above 15 seed
vortices) the dependence saturates at about $T_{\rm on}/T_{\rm c}
\leq 0.55$. } \label{PipeOnsetTempVsN}
\end{center}
\vspace{-8mm}
\end{figure}

In Figs.~\ref{PipeOnsetTemp} and \ref{PipeOnsetTempVsN} the
calculations from Fig.~\ref{VorReconTubeFlow} are carried further,
to check for a dependence of the onset temperature of turbulence
on the flow velocity $v_{\rm s}$ and on the number of seed
vortices $\mathcal{N}_{\rm i}$. In Fig.~\ref{PipeOnsetTemp} an
almost linear dependence is found for $T_{\rm on}$, as a function
of the velocity $v_{\rm s}$ which is imposed on the superfluid
component. This line represents the onset temperature of the
single-vortex instability: At and below this critical temperature
a state of sustained turbulence is observed to develop. At very
low velocities $(v_{\rm s} \lesssim 0.4\,$mm/s) a region is found
where no vortex instability and no dynamic generation of new
vortices occurs at any temperature down to below $0.3\,T_{\rm c}$.
This almost vertical boundary we associate with the critical
velocity of turbulence at vanishingly small mutual friction. In
Fig.~\ref{PipeOnsetTempVsN} in turn, we see that the onset
temperature depends on the initial number of seed vortices with
which the calculation is started: A roughly linear dependence
$T_{\rm on} (\mathcal{N}_{\rm i})$ is found.
Figs.~\ref{PipeOnsetTemp} and \ref{PipeOnsetTempVsN} are thus both
consistent with what we expect on the basis of the rotating
experiments (compare to Figs.~\ref{PrecursorSlope} and
\ref{BurstTime}). In our calculations on rotating flow these
dependences have not emerged with equal clarity,  owing to the
higher stability of evolving vortices in the rotating cylinder. In
contrast pipe-flow calculations, performed with the same
conventions and approximations, straightforwardly lead to the
expected relations.

\section{SUMMARY}

A quantized vortex is a topologically stable structure of the
superfluid order parameter field. In principle, it should be
possible to account for the appearance of every new vortex. In
$^4$He-II this has notoriously been a difficult task: Vortices
appear to emerge out of nowhere, without apparent systematics.
Various mechanisms have been proposed to explain their origin. In
$^3$He-B at temperatures below 2.3\,mK vortex formation processes
are instabilities and not thermally activated. Vortex formation is
here in better control, partly owing to the two or three orders of
magnitude larger vortex core diameter and the reduced influence
from surface roughness. Moreover, vortex formation can here be
examined as a function of a mutual friction dissipation which has
a strong, almost exponential temperature dependence around the
critical regime $\zeta \sim 1$.

With decreasing mutual friction the stability of vortices is
reduced and at $T \sim 0.6\,T_{\rm c}$ turbulence in the bulk
volume becomes possible. This hydrodynamic transition from laminar
to turbulent dynamics takes place at a somewhat higher temperature
than where an isolated evolving vortex might become unstable in
the rotating column. Before turbulence can be started from single
vortices evolving at low density, new expanding loops have to be
generated by a precursor mechanism. This happens via the
single-vortex instability for which the probability rapidly
increases with reducing friction. Thus the cascade process, the
single-vortex instability followed by a localized turbulent burst
in the bulk volume, becomes possible. Experimentally this is
observed as an abrupt change in the stability of the dynamics,
manifested as a sudden transition to turbulence within a narrow
temperature interval. The simplest means to investigate the
instability is to measure at constant rotation velocity the onset
temperature $T_{\rm on}$ of turbulence after the introduction of a
controlled number of seed vortices. Examples of such measurements
are shown in Figs.~\ref{RemOnsetTempLongCyl} and
\ref{TopBottomComparisonTon}, where $T_{\rm on}$ has been
determined in each of the four graphs. In Sec.~\ref{Measurements}
we have analyzed individual data points in such graphs, in order
to characterize the properties of the single-vortex instability.
It thereby turns out that in the onset temperature regime, $T
\approx T_{\rm on}$, a fraction of the transitions to turbulence
display prolonged precursory vortex formation at slow rate, before
the turbulent burst in the bulk sets in. At temperatures further
below the onset regime the instabilities proceed too rapidly to be
captured with our measuring techniques.

Our numerical simulation calculations in
Sec.~\ref{VortexFilamentCalculation} show that interactions of the
evolving seed vortex with the cylindrical wall in the presence of
rotating counterflow is the predominant source for new vortices in
the low density regime, before interactions between vortices in
the bulk become possible. In these calculations the walls are
represented with the boundary conditions of an ideal solid
surface. The experimental results suggest, however, that surface
properties or geometrical features do influence the onset
temperature, as seen in Fig.~\ref{BurstLocation}. More realistic
boundary conditions might therefore be needed and might reduce the
main disagreement between the rotating experiments and present
calculations, namely enhance the probability of wall reconnections
and the formation of new expanding loops in the calculations. In
contrast, our calculations on flow in a straight circular pipe
suggest that in this geometry wall interactions lead to the
generation of new vortices at higher temperatures and at lower
flow velocities than in rotation. Of these two types of flow,
rotation, which promotes polarization of vortices along the
rotation axis, is a more stable environment for dynamically
evolving vortices. This is in particular the case for a long
smooth-walled circular column with good alignment along the
rotation axis. The aligned circular column appears to be a most
stable special case \cite{Eltsov:2008} compared to, for instance,
a column which is inclined by a large amount with respect to the
rotation axis or one with square cross section \cite{Manchester}.
Similar differences in stability are known to apply for viscous
pipe flow with circular versus square cross section of the flow
channel \cite{Mullin}.

Finally we note that the single-vortex instability starts in the
rotating column a sequence of events, which have been described in
a recent review \cite{Eltsov:2008}. Of central interest here is
the propagation of polarized vortices along the rotating circular
column and how this is changed by the increasing turbulent
influence with decreasing temperature \cite{Front}. These
phenomena elucidate superfluid turbulence in the $T \rightarrow 0$
limit, which recently has been studied also in two other types of
measurements, by monitoring the decay of homogeneous isotropic
turbulence, after the external pumping is switched off
\cite{Manchester,Lancaster}.

\textbf{Acknowledgments:}---This work is supported by the Academy
of Finland (grants 213496, 124616, and 114887) and by ULTI
research visits (EU Transnational Access Programme FP6, contract
RITA-CT-2003-505313).

\end{document}